\documentclass[epj]{svjour}
\usepackage{graphics}
\usepackage{amssymb}
\newcommand{\be}{\begin{equation}}
\newcommand{\ee}{\end{equation}}
\newcommand{\bd}{\begin{displaymath}}
\newcommand{\ed}{\end{displaymath}}
\newcommand{\BE}{\begin{eqnarray}}
\newcommand{\EE}{\end{eqnarray}}

\newcommand{\bra}{\left\langle}
\newcommand{\ket}{\right\rangle}

\newcommand{\sgn}{{\rm sgn}}
\newcommand{\erf}{{\rm erf}}
\newcommand{\id}{{\rm 1\!I}}

\newcommand{\bq}{\ensuremath{\mathbf{q}}}

\newcommand{\Rbo}{{\mbox{\boldmath $R$}}}

\newcommand{\xibo}{{\mbox{\boldmath $\xi$}}}

\newcommand{\omegabo}{{\mbox{\boldmath $\omega$}}}
\newcommand{\boldpsi}{{\mbox{\boldmath $\psi$}}}

\newcommand{\Omegabo}{{\mbox{\boldmath $\Omega$}}}

\newcommand{\tp}{t^\prime}
\newcommand{\td}{t^{\prime\prime}}

\newcommand{\rhobar}{\overline{\rho}}
\begin{document}
\title{Strategy correlations and timing of adaptation in Minority Games}
\author{Tobias Galla\dag\ddag$\|$ and David
Sherrington\dag}% etc
%}                     % Do not remove
\institute{\dag The Rudolf Peierls Centre for Theoretical Physics, University of Oxford, 1 Keble Road, Oxford OX1 3NP, UK\\
\ddag\ International Center for Theoretical Physics, Strada Costiera 11, 34014 Trieste, Italy\\
$\|$ Istituto Nazionale per la Fisica della Materia (INFM), Trieste-SISSA Unit, V. Beirut 2-4, 34014 Trieste, Italy}
\date{\today}
\abstract{
We study the role of strategy correlations and timing of adaptation
for the dynamics of Minority Games, both simulationally and
analytically. Using the exact generating functional approach \`a la De
Dominicis we compute the phase diagram and the behaviour of batch and
on-line games with correlated strategies, complementing exisiting
replica studies of their statics. It is shown that the timing of
adaptation can be relevant; while conventional games with uncorrelated
strategies are nearly insensitive to the choice of on-line versus
batch learning, we find qualitative differences when anti-correlations
are present in the strategy assignments. The available standard
approximations for the volatility in terms of persistent order
parameters in the stationary ergodic states become unreliable in batch
games under such circumstances. We then comment on the role of
oscillations and the relation to the breakdown of ergodicity. Finally,
it is discussed how the generating functional formalism can be used to
study mixed populations of so-called `producers' and `speculators' in
the context of the batch minority games.}

 % end of PACS codes
 %end of abstract
%
\maketitle
\section{Introduction}
The collective behaviour of interacting heterogeneous and adaptive
agents has attracted a substantial amount of attention in the
statistical physics communtity in recent years. The general aim is to
understand how complex macroscopic co-operation can emerge from the
underlying microscopic laws that govern the behaviour of the
individual agents. The minority game (MG) \cite{ChalZhan97} is
probably one of the most studied models in this context. It describes
an ensemble of traders who at each time step receive a piece of public
information and react by either buying or selling. Learning from past
experience, the aim of the individual agent is to be in the minority
at each round of the game, i.e. to buy when most of the traders are
selling and vice versa. To take their trading decisions the agents
employ strategies assigned randomly at the start of the game and then
kept fixed. These effectively act as look-up tables providing a map
from the observed public information onto a binary trading
decision. At each time step the individual agent employs the strategy
in his or her pool he or she believes will maximise their potential
payoff. In order to measure the performance of their strategies they
allocate virtual points to each of their strategies, increasing the
score of strategies that would have yielded a correct minority
decision.

Thanks to the simplicity of the microscopic equations of motion and
because of analogies to models of spin-glasses and neural networks,
analytical progress can be made using both static and dynamical
methods of statistical mechanics. It has been found that the relevant
control parameter for MGs is the ratio $\alpha=P/N$ of the number $P$ of
possible values for the external information over the number of players $N$.  A
phase transition between an ergodic phase at high $\alpha$ and
non-ergodic states below a critical value $\alpha_c$ has been
identified and analytical expressions for the macroscopic order
parameters characterising the different states have been obtained
\cite{ChalMarsZhan00,ChalMarsZecc00,MarsChalZecc00,GarrMoroSher00,HeimCool01,MarsChal01}. While some
order parameters in the ergodic phase can be computed in the
thermodynamic limit $N\to\infty$ without making any approximations,
the dynamics of the game in its non-ergodic phase at low values of
$\alpha$ is not yet fully understood. Furthermore the calculation of a
key observable of the MG, the magnitude of the overall market
fluctuations or so-called volatility, is up to now restriced to
approximations both in dynamical approaches and in static replica
analyses \cite{HeimCool01,ChalMarsZecc00}.

In the original setup of the MG the strategies assigned to a given
agent were completely uncorrelated and the information fed to the
agents was based on the real history of the market dynamics
\cite{ChalZhan97,SaviManuRiol99}. Also, updates of the virtual scores
of the agents' strategies were performed after each round of the game,
and their relative success re-assessed before each trading decision,
i.e. agents could switch strategies between any two successive rounds
of the game.  Such variants of the MG with score updates at every
round of the game are generally referred to as `on-line' games. It was
realised that no qualitative changes in the overall behaviour occured
when the stream of information was replaced by random inputs at each
time step \cite{Cava99}. Hereafter we shall therefore distinguish
between on-line games with `real' or `random' market history,
respectively. It is technically much easier to deal with the case of
`random' history so henceforth we concentrate on this version (see
however the recent work \cite{Cool04} for a generating functional
analysis of MGs with real market histories).  Early studies showed
only slight differences between on-line MGs with uncorrelated
strategies and corresponding games in which the agents effectively
sample a large set of information patterns before updating the scores
of their strategies. This amounts to replacing the inflowing
information by an averaging procedure over all possible values of the
pieces of information to yield an effective interaction between the
agents; such models are known as `batch' minority games. On the
technical level, using this effective interaction, the dynamics of
batch MGs are much easier to study than their on-line counterparts,
and hence most current dynamical studies are concerned with batch
versions of the MG
\cite{HeimCool01,HeimDeMa01,DeMaGiarMose03,DeMa03,GallCoolSher03,Gall05}. Also, here we restrict ourselves to the case of deterministic
decision making; a stochastic extension is straightforward to consider
\cite{CoolHeimSher01}, but complicates the detailed mathematics of
the analysis. Reviews on the MG can be found in \cite{Coolenreview,Moro04}; note also the recently published textbooks \cite{MGbook1,MGbook2}.

In this paper we extend the generating functional analysis of batch
and on-line games with uncorrelated strategies as presented in
\cite{HeimCool01,CoolHeim01} to the case of correlated and anti-correlated
strategies. The purpose is twofold, firstly we provide an analysis of
the dynamics of such MGs with so-called `diversified' strategies and
complement the study of their statics previously presented by other
authors in \cite{ChalMarsZhan00}. Secondly, we find that in the
presence of anti-correlated strategies significant differences in the
global behaviour of batch and on-line MGs are present, i.e. that the
the overall performance of the system strongly depends on the timing
of adapation of the agents. This effect is small in the case of
conventional MGs with uncorrelated strategies, but becomes magnified
as the degree of anti-correlation in the strategy assignments is
increased. We also demonstrate that some approximations which
reproduce the the volatility of batch MGs with uncorrelated strategies
with good accuracy, and are hence often considered to be standard, become
inaccurate in games with highly anti-correlated stategies. We then
discuss the effects of strategy correlations on the sensitivity to
initial conditions and comment on the relation between the global
oscillations and the breakdown of ergodicity in batch and on-line MG
with different strategy correlations. As a by-product of the
generating functional analysis of the dynamics of games with general
distributions of correlation parameters we finally study the case of
bi-modal distributions, corresponding to a mixed population of
so-called `speculators' and `producers'

\section{Model Definitions}
%%%%%%%%%%%%%%%%%%%%%%%%%%%%%%%%%%%%%%%%%%%%%%%%%%%%%%%%%%%%%%%%%%%%%%%%%%%%%%
The MG describes the decision making dynamics of $N$ interacting
agents, $i=1,\dots,N$. At each round $\ell$ of the game all agents are
given the same external information $I(\ell)$ taken from a set of $P$
elements. They utilise this information to determine an action by
choosing from one of a personal set of $S$ strategies whose operation
on the information yields a decision. Simulations have shown that
restriction to $S=2$ strategies per agent yields characteristic
behaviour. It also simplifies the analysis. Hence we restrict the
further discussion here to this case. We also simplify the choices of
the space of information and character of the decisions. Hence here we take
each agent to hold two $P$-dimensional strategy vectors
$\Rbo_{ia}=(R_{ia}^1,\dots,R_{ia}^{P})$, $a=\pm 1$, with each
component $R_{ia}^\mu$ chosen randomly from the set $\{-1,1\}$ at the
beginning of the game and thereafter fixed and the information to be
given by an integer $\mu(\ell)$ chosen randomly and independently at
each step $\ell$ from the set $\{1,\dots,P=\alpha N\}$.  Then the strategies
act as look up tables for deciding `trading' action, yielding
output $R^{\mu(\ell)}_{ia}$. If player $i$ decides to use strategy
$a^*_i(\ell)$ at round $\ell$ of the game his or her trading action
will be $b_i(t)=R_{i{a_i^*}(\ell)}^{\mu(\ell)}$. The re-scaled total
bid at stage $\ell$ is then defined as $A(\ell)=N^{-1/2}\sum_i
b_i(\ell)$. In order to decide which of their two strategies to use at
any given time each players keeps a record of the relative performance
of his or her strategies. A payoff value $p_{ia}(\ell)$ is assigned to
each strategy and updated every $M$ time steps according to
\begin{equation}\label{update}
p_{ia}(\ell+M)=p_{ia}(\ell)-\frac{\Gamma}{\sqrt{N}}\sum_{\ell^\prime=\ell}^{\ell+M-1}
R_{ia}^{\mu(\ell^\prime)}A(\ell^\prime).
\end{equation}
In between the updates the scores $p_{ia}$ are kept
constant. $\Gamma>0$ is a learning rate and of order ${\cal O}(N^0)$,
introduced for convenience \cite{CoolHeim01}. Note that the minus sign
in the update prescription ensures that strategies which predict
correct minority decisions are rewarded. At each round player $i$ then
plays the strategy in his or her arsenal with the highest relative
score, i.e. $a^*_i(\ell)=\sgn[q_i(\ell)]$, where
$q_i(\ell)=\frac{1}{2}(p_{i,1}(\ell)-p_{i,-1}(\ell))$ is the point
difference of player $i$'s strategies at time $\ell$. Upon
introduction of $\omegabo_i=(\Rbo_{i,1}+\Rbo_{i,-1})/2$ and
$\xibo_i=(\Rbo_{i,1}-\Rbo_{i,-1})/2$ the update rule (\ref{update})
can be compactified to
\begin{eqnarray}
q_i(\ell+M)&=&q_i(\ell)-\frac{\Gamma}{\sqrt{N}}\sum_{\ell^\prime=\ell}^{\ell+M-1}\xi_i^{\mu(\ell^\prime)}\nonumber\\
&&\hspace{-1.5cm}\times\bigg\{N^{-1/2}\sum_j
\left(\omega_j^{\mu(\ell^\prime)}+\xi_j^{\mu(\ell^\prime)}s_j(\ell^\prime)
\right)\bigg\}.\label{eq:updatem}
\end{eqnarray}

Here we have introduced the shorthand notation $s_i(\ell)=\sgn[q_i(\ell)]$, we will also abbreviate $\Omegabo=N^{-1/2}\sum_j \omegabo_j$. The case
$M=1$ (where strategy scores are updated at every time step) is
referred to as the on-line model. For $M\geq{\cal O}(\alpha N)$ one
expects (and simulations confirm) the behaviour of the system to be
the same as that for the so-called batch version in which the sum over
the actual values $\mu(\ell^\prime)$ is replaced by an average over
the set $\mu\in\{1,\dots,\alpha N\}$:
\begin{eqnarray}\label{eq:batchupdate1}
q_i(t+1)&=&q_i(t)-\frac{\Gamma}{N}\sum_{\mu=1}^{\alpha N}\xi_i^\mu\big\{\sum_j\left(\omega_j^\mu+\xi_j^\mu s_j(t)\right)\big\}.
\end{eqnarray}
Note that an appropriate re-scaling of time is implied, as one batch
time step corresponds to ${\cal O}(\alpha N)$ on-line
steps. Introducting the notation
$J_{ij}=\Gamma\xibo_i\cdot\xibo_j/N$ and
$h_i=\Gamma\xibo_i\cdot\Omegabo/\sqrt{N}$ the batch update
rule (\ref{eq:batchupdate1}) can be written compactly as
\begin{equation}\label{eq:batchupdatejh}
q_i(t+1)=q_i(t)-h_i-\sum_j J_{ij}s_j(t).
\end{equation}
In the original game the strategies are chosen without correlations between the agents and also independently within the set of strategies of a fixed agent $i$, i.e. one has
\be
\overline{R_{ia}^\mu R_{jb}^\nu}=\delta_{ij}\delta_{ab}\delta_{\mu\nu},
\ee
where $\overline{\cdots}$ denotes an average over the disorder,
i.e. over the space of strategy assignments. In this paper, we will
consider cases of correlated strategy vectors $\Rbo_{i,-1}$ and
$\Rbo_{i,1}$ of a fixed agent $i$. Specifically we generalize the standard case to a situation in which $R_{i1}^\mu=R_{i2}^\mu$
with probability $\rho_i\in [0,1]$ for any given $i$ and $\mu$ (and
$R_{i1}^\mu\neq R_{i2}^\mu$ with probability $1-\rho_i$, recall that the $R_{ia}^\mu$ take only values $-1$ and $1$). The joint
probability distribution of $R_{i1}^\mu$ and $R_{i2}^\mu$ is then given by
\BE
P(R_{i1}^\mu=x,R_{i2}^\mu=y)&=&\frac{\rho_i}{2}\left(\delta_{x,-1}\delta_{y,-1}
+\delta_{x,1}\delta_{y,1}\right)\nonumber \\
&&\hspace{-1cm}+\frac{1-\rho_i}{2}\left(\delta_{x,-1}\delta_{y,1}
+\delta_{x,1}\delta_{y,-1}\right).
\EE
The standard situation of independent strategies is covered as the
special case $\rho_i=1/2$ for all players $i$. On the other hand, if
$\rho_i=1$, then player $i$ holds two identical strategy vectors,
whereas for $\rho_i=0$ he or she has two opposite strategies,
$\Rbo_{i,-1}=-\Rbo_{i,1}$. Allowing the correlation parameter $\rho_i$
to depend on $i$ adds another layer of heterogeneity to the ensemble
of agents: not only are their strategies chosen at random and hence
are heterogeneous across the group of $N$ players, but also the
probability distribution from which they are drawn may be different
for different agents. We will assume that each $\rho_i$ is randomly
and independently drawn from a distribution $P(\rho)$, so that, for
example, $P(\rho)=\delta(\rho-1/2)$ corresponds to the standard game,
with $\rho_i=1/2$ for all $i$. Some of our results presented below
have been summarized in the short paper \cite{SherGall03}. The statics
of MGs with correlated strategies were first studied using replica
methods in
\cite{ChalMarsZhan00}. Numerical results are also found in
\cite{GarrMoroSherr01}, and in \cite{YipHuiLoJohn03}, where an effective correlation between the assigned strategies was introduced upon biasing the $\{R_{ia}^\mu\}$ towards one of the binary entries.

Let us finally, in this section, introduce the volatility, $\sigma^2$,
of the market. It describes the variance of the total
re-scaled market bid $A$, and can be defined as the following long-time average:
\be
\sigma^2=\lim_{\tau\to\infty}\tau^{-1}\sum_{\ell\leq \tau}A(\ell)^2.
\ee
In on-line models the relevant average over the stochasticity of the information is to be performed, as $A(\ell)$ depends on both the score valuations $\{q_i(\ell)\}$ and the value of the external information $\mu(\ell)$.  In deterministic batch games this average is replaced by one over $\mu$
\be
\sigma^2=\lim_{\tau\to\infty}\tau^{-1}\sum_{t\leq\tau}P^{-1}\sum_{\mu=1}^P (A^\mu(t))^2,
\ee
where $A^\mu(t)=\Omega^\mu+N^{-1/2}\sum_j\xi_j^\mu s_j(t)$.
 The volatility is a measure for the global
efficiency of the market; if $\sigma^2=0$ supply and demand are always
matched ($A=0$) and trading is fully efficient. Non-zero volatilities
however indicate mismatches between the numbers of buyers and sellers,
and hence imply inefficient markets. For the mathematical analysis of batch MGs it is also convenient to introduce the volatility matrix $\Xi$ \cite{HeimCool01}
\be
\Xi_{t\tp}=P^{-1}\sum_{\mu=1}^P A^\mu(t)A^\mu(\tp).
\ee
Note that $\sigma^2=\lim_{\tau\to\infty}\tau^{-1}\sum_{t\leq\tau}\Xi_{tt}$, and that stochastic trading with $s_i(t)=\pm 1$ taken randomly and independently at any time step $t$ would result in $\Xi_{t\tp}=\frac{1}{2}(1+\delta_{t\tp})$ in the thermodynamic limit. In particular $\sigma^2=1$ is the so-called random trading limit.

%%%%%%%%%%%%%%%%%%%%%%%%%%%%%%%%%%%%%%%%%%%%%%%%%%%%%%%%%%%%%%%%%%%%%%%%%%%%%%
\section{Generating functional and effective single-agent process for
the batch game}\label{sec:gfa}
We will now turn to the generating
functional analysis of the batch process (\ref{eq:batchupdate1}). The calculation is an extension of the analysis
of the standard game presented in \cite{HeimCool01}.
\subsection{Generating functional and disorder average}
The moment generating functional corresponding to the batch process of
Eq. (\ref{eq:batchupdate1}) reads
\BE
Z[\boldpsi]&=&\int d\bq\, p_0(q(0)) \exp\left(i\sum_{it}\psi_i(t)s_i(t)\right)
\nonumber\\
&&\hspace{-1.6cm}\prod_{it} \delta\left(q_i(t+1)-q_i(t)+\sum_j
 J_{ij}s_j(t)+h_i-\theta_i(t)\right),
\EE
where we have introduced the usual source term containing the field
variables $\{\psi_i(t)\}$. The $\{\theta_i(t)\}$ are additional perturbation fields, introduced to generate response functions, and $p_0(q(0))$ denotes the probability distribution of the initial score differences $q(0)=(q_1(0),\dots,q_N(0))$ from which the dynamics is started. 

The further analysis is standard and follows the lines of
\cite{HeimCool01}. The average over the strategy assignments can be
performed exactly in the thermodynamic limit $N\to\infty$, leading to
a (quenched-strategy-) disorder averaged generating functional $\overline{Z[\boldpsi]}$,
from which all dynamical observables can be computed upon taking
derivatives with respect to the $\{\psi_i(t)\}$ and
$\{\theta_i(t)\}$, finally taking the limit in which these generating fields go to zero. The relevant macroscopic order parameters
of the resulting theory are the weighted disorder-averaged correlation and
response functions of the original $N$-particle problem:
\begin{eqnarray}
C_{t\tp}&=&\lim_{N\to\infty}\frac{1}{N}\sum_i \frac{1-\rho_i}{1-\rhobar}\overline{\bra s_i(t) s_i(t^\prime)\ket}, \label{eq:origC} \\
G_{t\tp}&=&\lim_{N\to\infty}\frac{1}{N}\sum_i \frac{1-\rho_i}{1-\rhobar}\frac{\partial \overline{\bra s_i(t)\ket}}{\partial\theta_i(t^\prime)}\label{eq:origG},
\end{eqnarray}
where $\bra\dots\ket$ denotes an average over initial conditions (as
specified by the distribution $p_0(q(0))$. We use the notation
$\overline{\rho}=\int d\rho P(\rho)\rho$ for the first moment
of the distribution $P(\rho)$ of correlation parameters\footnote{Note
that the overbar on $\rho$ has a different meaning from the other
overbars, which indicate averaging over the quenched disorder
choices.}. While the numerator in the unusual pre-factors
$(1-\rho_i)/(1-\rhobar)$ is a direct consequence of the averaging
procedure over the assignments of correlated strategies, the
denominator is chosen to ensure that the equal time correlation
function is equal to one, $C_{tt}=1$\footnote{We assume here that
$\rhobar<1$, and exclude the case $\rhobar=1$. The latter case
corresponds to a situation in which all agents (except for a
non-extensive number) have $\rho_i=1$, and hence hold two identical
strategies. Such a game trivially leads to a volatility $\sigma^2=1$,
corresponding to the random trading limit, and exhibits no phase
transition.}. For later convenience we shall take the inverse
characteristic time-scale $\Gamma$ in the couplings $J_{ij}$ and the
fields $h_i$ to be given by $\Gamma=1/(1-\rhobar)$.

In an extention of the procedure for uncorrelated agents, by introducing
auxiliary macroscopic functions relateable to the correlation and response
functions, performing summations over the microscopic variables and
utilising extremal dominance in the limit $N\to\infty$, the
computation results in an equivalence of the original Markovian coupled
$N$-particle dynamics to an ensemble of non-Markovian single agent problems,
explicitly involving the correlation and response functions $C$ and $G$ and
subject to self-consistently determined coloured noise.

Explicitly, the equivalence is to an ensemble of effective single agents with characteristic labels $\rho$ subject to stochastic dynamics
\BE\label{eq:singleagent}
q_\rho(t+1)&=&q_\rho(t)+\theta_\rho(t)-\frac{1-\rho}{1-\rhobar}
\alpha\sum_{t^\prime\leq t}(\id+G)^{-1}_{t\tp}s_\rho(t^\prime)\nonumber\\
&&+\sqrt{\alpha\frac{1-\rho}{1-\rhobar}}\eta_\rho(t);\qquad s_\rho(t)=\sgn[q_\rho(t)]
\EE
with fractional population $P(\rho)$ and with $\{\eta_\rho(t)\}$ Gaussian coloured noise of zero average and covariance
\begin{equation}\label{eq:correlcovariance}
\Lambda_{t\tp}\equiv\bra\eta_\rho(t)\eta_\rho(t^\prime)\ket_{\eta_\rho}=[(\id+G)^{-1}D(\id+G^T)^{-1}]_{t\tp},
\end{equation}
where $\id_{t\tp}=\delta_{t\tp}$ and $C$, $G$ and $D$ ($D_{t\tp}=\frac{\overline{\rho}}{1-\rhobar}+C_{t\tp}$ for all $t,\tp$) are determined self-consistently across the ensemble\footnote{A similar situation involving an ensemble of single agent processes
has been encountered in \cite{DeMa03}, where the author discusses a
population of agents trading with different frequencies.}.  In the
derivation of (\ref{eq:singleagent}) we have assumed that the
perturbation fields $\theta_i(t)$ for players $i$ with correlation
parameter $\rho_i=\rho$ are all identical,
$\theta_i(t)=\theta_\rho(t)$, and that the initial values $q_i(0)$ for
agents $i\in\{j=1,\dots,N|\rho_j=\rho\}$ in the original $N$-particle
dynamics are all drawn independently from the same distribution
$p_{0\rho}(q_i(0))$, i.e. that the distribution of initial values
factorizes over agents with the same correlation parameter $\rho$.

Note that the different single particle noises $\{\eta_\rho\}$ (for
different values of $\rho$) all have the same covariance matrix
$\Lambda_{t\tp}$, with no explicit dependence on $\rho$, but only an
implicit one on the distribution $P(\rho)$ through the correlation and
response matrices $C$ and $G$. Nevertheless we will keep the subscript
$\rho$ in $\eta_\rho$ to distinguish the noise contributions to the
effective single-agent processes for different values of $\rho$. The
correlation and response functions $C_{t\tp}$ and $G_{t\tp}$ are then
to be computed self-consistently as two-fold averages over (i) the
measure $\bra\dots|\rho\ket_*$ generated by the realizations of the
process $\{q_\rho(t)\}$:
\BE\label{eq:starmeasure} <f[q_\rho]|\rho>_* &=& \int
Dq_\rho D\hat q_\rho p_0(q_\rho(0)) f[q_\rho] \nonumber \\
&&\hspace{-2cm}\times
\exp\left(i\sum_{t}\hat q_\rho(t)(q_\rho(t+1)-q_\rho(t)-\theta_\rho(t))\right) \nonumber
\\ 
&& \hspace{-2cm}\left. \times
\exp\left(i\alpha\frac{1-\rho}{1-\rhobar}\sum_{tt^\prime}\hat
q_\rho(t)(\id+G)^{-1}_{t\tp}s_\rho(t^\prime)\right.\right.\nonumber\\
&&\hspace{-2cm}
\left.-\frac{1}{2}\alpha\frac{1-\rho}{1-\rhobar}\sum_{tt^\prime}\hat
q_\rho(t)[(\id+G)^{-1} D (\id+G^T)^{-1}]_{t\tp} \hat
q_\rho(t^\prime)\right),\nonumber\\
\end{eqnarray} 
and (ii) over the distribution $P(\rho)$ of correlation parameters. In detail one has
\BE
C_{t\tp}&=&\int d\rho P(\rho) \frac{1-\rho}{1-\rhobar} <s_\rho(t)s_\rho(t^\prime)|\rho>_*
\nonumber\\
&\equiv& \int d\rho P(\rho) \frac{1-\rho}{1-\rhobar} C_\rho(t,\tp),\label{eq:scc}\\
G_{t\tp}&=&\int d\rho P(\rho) \frac{1-\rho}{1-\rhobar} \frac{\partial}{\partial\theta_\rho(t^\prime)}<s_\rho(t)|\rho>_*\nonumber \\
&\equiv& \int d\rho P(\rho) \frac{1-\rho}{1-\rhobar} G_\rho(t,\tp)\label{eq:scg},
\EE
so that the different processes (\ref{eq:singleagent}) (with different
values of $\rho$) are effectively coupled through the matrices $C$ and
$G$, which appear in (\ref{eq:singleagent}) and in the noise
correlator (\ref{eq:correlcovariance}) and which at the same time are
(weighted) averages over the whole ensemble of representative agent
processes. The original $N$-agent dynamics and the self-consistent
single agent problem are equivalent in the thermodynamic limit in the
sense that disorder-averaged observables in the original problem (such
as the correlation and response functions (\ref{eq:origC},
\ref{eq:origG})) are identical to the corresponding averages obtained from the ensemble of effective single-agent processes (Eqs. (\ref{eq:scc}, \ref{eq:scg}) for the correlation and response functions). This equivalence extends to other macroscopic observables,
including the probability for a given agent to `freeze' (i.e. to use
only one of his or her two strategies) in the long-run. We will use
the resulting `fraction of frozen' agents below in the further
analysis of the effective agent dynamics.

Due to the presence of the coloured noise $\{\eta_\rho(t)\}$ and the
retarded self-interaction term in Eq. (\ref{eq:singleagent}) a direct
solution of the self-consistent problem defined by
(\ref{eq:singleagent}-\ref{eq:scg}) is in general impossible beyond
the first few time steps. An analysis of possible ergodic
time-translation invariant stationary states is however feasible, and
will be presented below. Alternatively one can resort to a numerical
iteration of the representative agent problem using a method first
proposed in \cite{EissOppe92}. This effective Monte-Carlo integration
of the single agent problem allows one to determine the correlation
and response matrices $C$ and $G$ to arbitrary precision without
finite-size effects, but becomes more and more costly computationally
as the number of time steps $t$ is increased (due to required
inversions of matrices of size $t\times t$ at time step $t$). 

\begin{figure}
\resizebox{0.5\textwidth}{!}{%
  \includegraphics{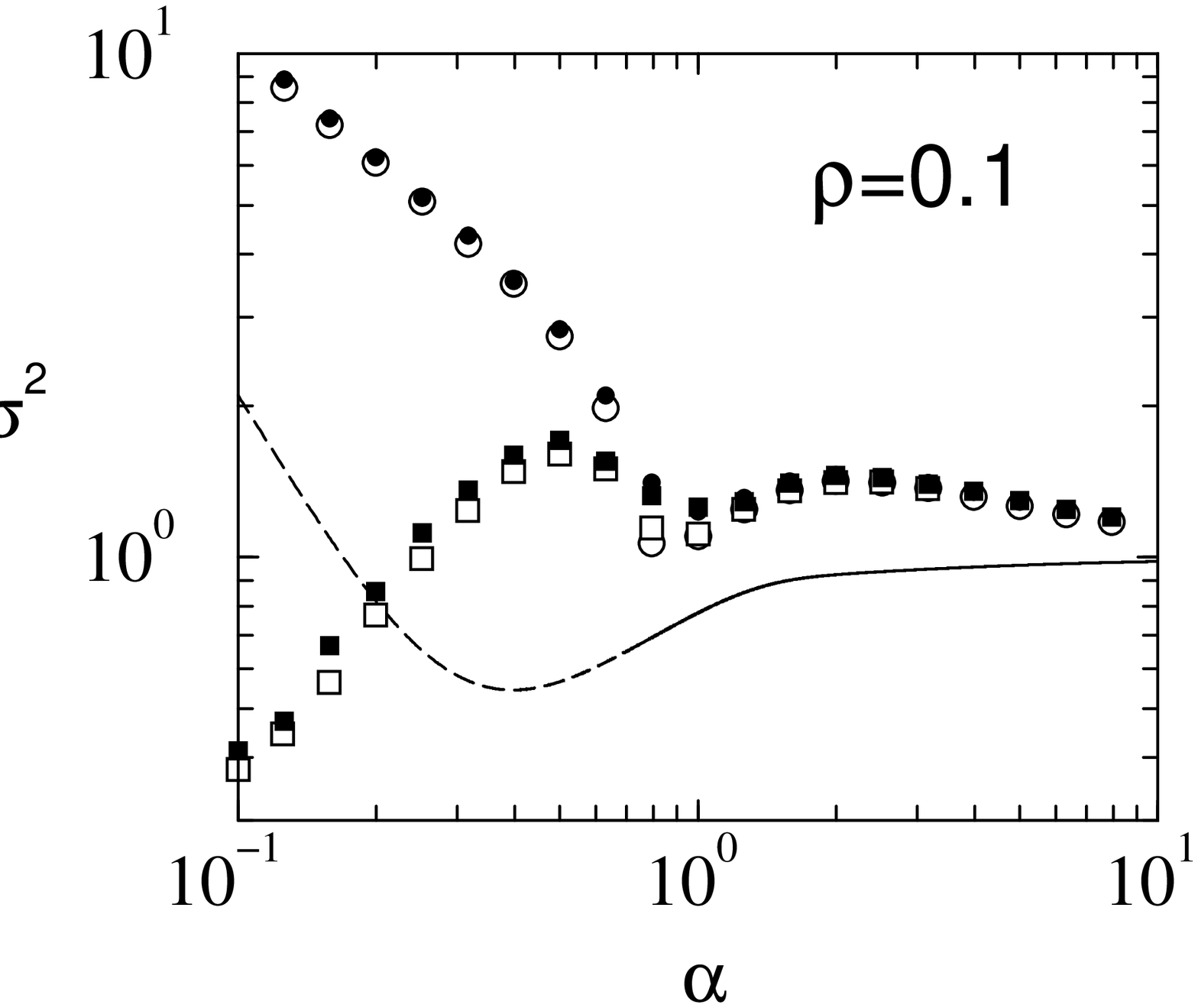} ~~~~
  \includegraphics{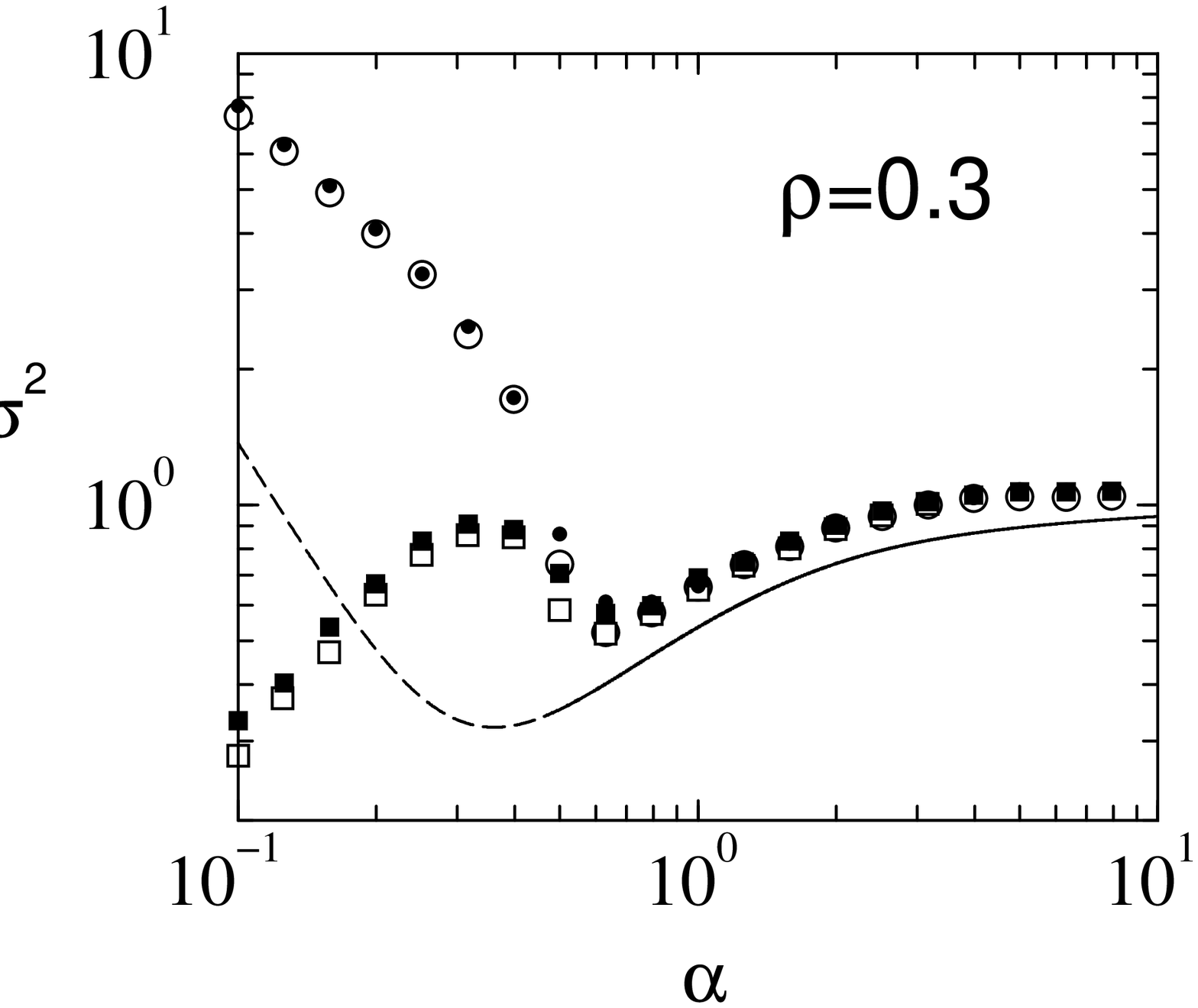}}\\
\resizebox{0.5\textwidth}{!}{%
 \includegraphics{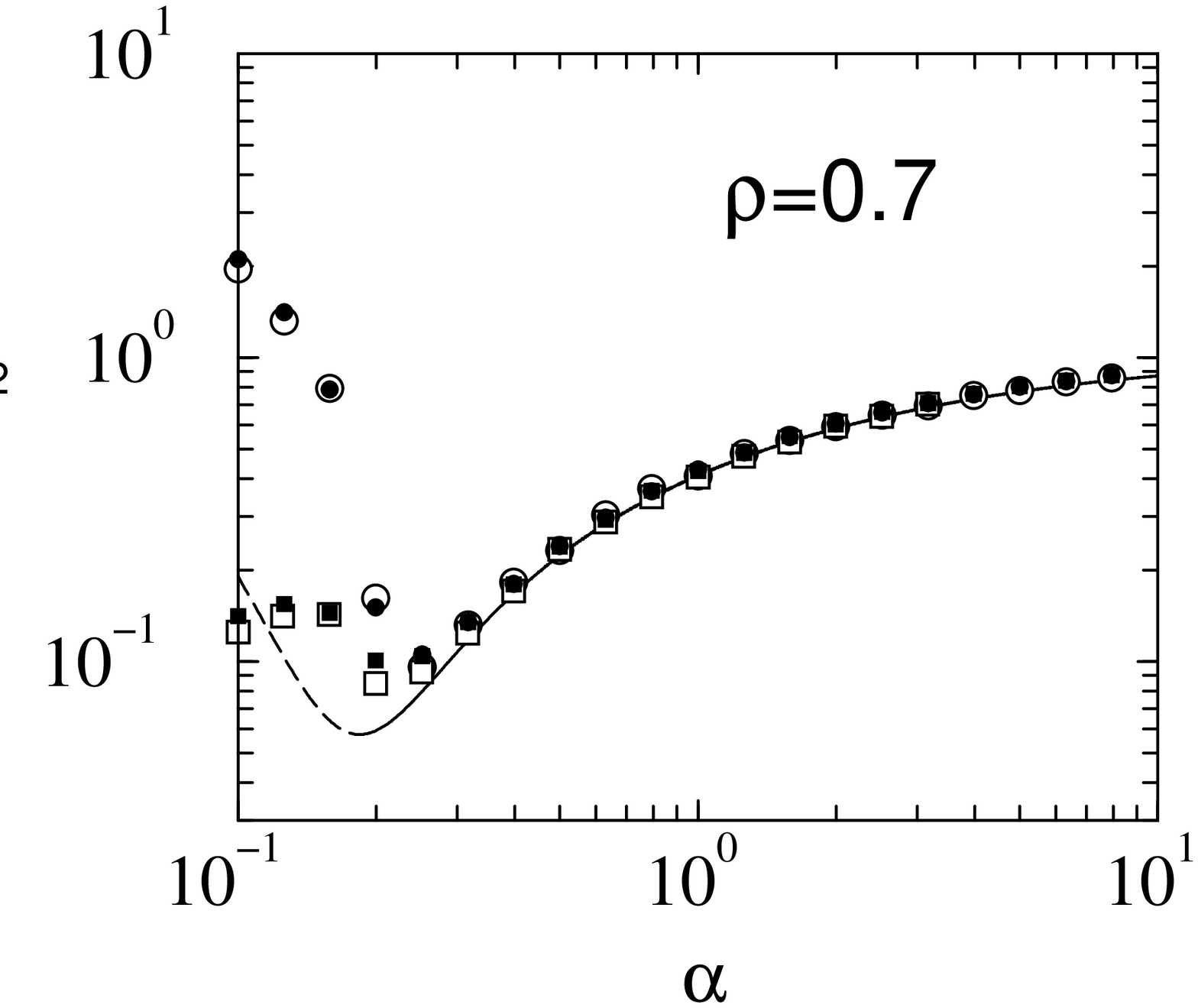} ~~~~
  \includegraphics{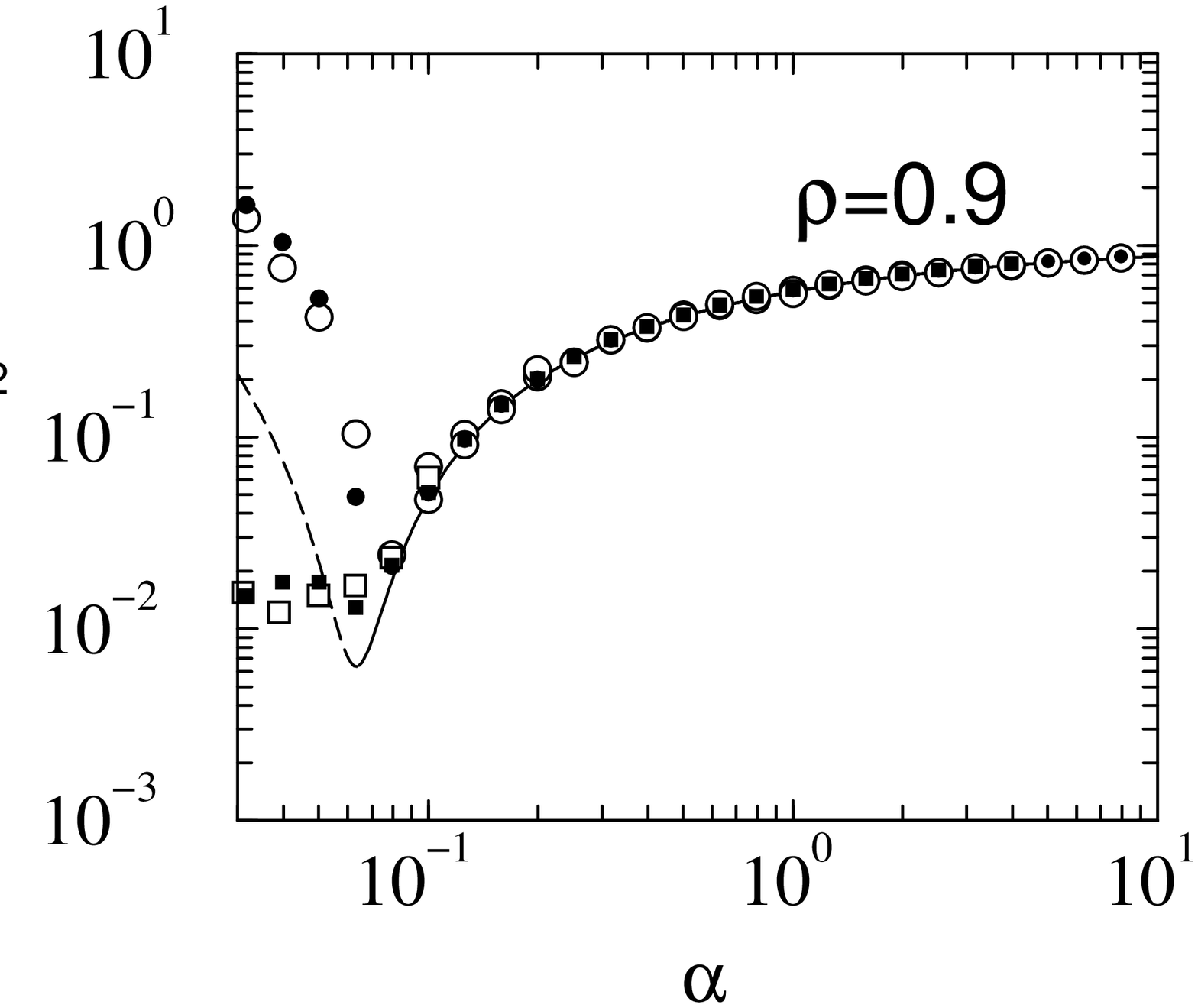}
}
\vspace{1em}
\caption{\label{vol_batch_rho} Volatility as a function
of $\alpha$ for different values of $\rho$ for the batch game
($\rho_i=\rho\,\forall i$). Open symbols are from simulations of the batch
process with $N=1000$ agents, averaged over $10$ samples of the
disorder, and run for $500$ batch time steps. Filled markers are
results from a direct iteration of the single-particle problem using
the method of \cite{EissOppe92} ($5\times 10^4$ realisations of the
single-agent process are generated, and the iteration is performed
for $100$ time steps). Circles: {\em tabula rasa start} ($q_i(0)=0$),
squares: biased starts ($|q_i(0)|=1$). Solid lines are the approximation of
(\ref{eq:vol_analytical}) for the ergodic phase and are continued as
dashed lines into the non-ergodic phase (where the ergodic theory is
longer valid). }
\end{figure}

This direct-iteration procedure does not require any further
assumptions on the properties of the dynamical order parameters, and
can be carried out for all values of $\alpha$ in both the ergodic and
non-ergodic regimes of the game.  In Fig. \ref{vol_batch_rho} we
display results for the volatility as a function of $\alpha$ for
unimodal distributions of the correlation parameter, $\rho_i=\rho$ for
all $i$. We find good agreement between the data obtained from a
direct simulation of the batch process (open symbols in
Fig. \ref{vol_batch_rho}) and from an iteration of the effective
single-agent process (solid markers), respectively. This confirms the
validity of the analytical theory derived in this section for general
$P(\rho)$ in the special case of unimodal distributions of strategy
correlations.

We will now proceed to analyse the single-effective agent process
further and to compute persistent order parameters in the ergodic
stationary state, as well as the phase diagrams of batch MGs with correlated strategies.
\subsection{Persistent order parameters in the ergodic stationary state}
In order to analyse the ergodic state in the regime of large $\alpha$
we will make the following assumptions: (i) the system becomes
stationary in the long-time limit, i.e. we will assume temporal translation
invariance, $\lim_{t\to\infty}C_{t+\tau,t}=C(\tau)$ and
$\lim_{t\to\infty}G_{t+\tau,t}=G(\tau)$, (ii) the absence of anomalous
response, i.e. we will assume that the integrated response
$\chi=\lim_{t\to\infty} \sum_{\tau\leq t}G(\tau)$ remains finite and
(iii) weak long-term memory, i.e. $\lim_{t\to\infty} G_{t\tp}=0$ for
all finite $\tp$.

The integrated response $\chi$ along with the persistent part of the
correlation function, $c=\lim_{t\to\infty}t^{-1}\sum_{\tau\leq t} C(\tau)$,
will be the order parameters characterising the ergodic stationary
states of the game. The further analysis hence proceeds by formulating
closed equations for these persistent order parameters.  

Following
\cite{HeimCool01} we introduce the re-scaled quantity $\tilde
q_\rho(t)=q_\rho(t)/t$, and find from (\ref{eq:singleagent})
\BE
\tilde q_\rho(t)&=&\frac{1}{t}q_\rho(0)
-
\frac{1-\rho}{1-\rhobar}\frac{\alpha}{t}\sum_{\tp<t}\sum_{\td}(\id+G)^{-1}_{\tp\td}s_\rho(\td)\nonumber\\
&&+\theta_\rho+\frac{\sqrt{\alpha}}{t}\sqrt{\frac{1-\rho}{1-\bar\rho}}\sum_{\tp<t}\eta_\rho(\tp)
\EE
Here, we have assumed that $\theta_\rho(t)=\theta_\rho$ for all $t$,
so that $\theta_\rho$ is a static perturbation. Now, upon taking the
limit $t\to\infty$ and writing $\tilde q_\rho=\lim_{t\to\infty}\tilde
q_\rho(t)$, we find
\be\label{eq:timeaverage}
\tilde q_\rho = -\alpha\frac{1-\rho}{1-\rhobar}\frac{s_\rho}{1+\chi}+\theta_\rho+\sqrt{\alpha\frac{1-\rho}{1-\rhobar}}\eta_\rho,
\ee
where $s_\rho=\lim_{t\to\infty} t^{-1}\sum_{\tau\leq t}\sgn[q_\rho(\tau)]$ and
\\$\eta_\rho=\lim_{t\to\infty} t^{-1}\sum_{\tau\leq t}\eta_\rho(\tau)$. Note that
$\tilde q_\rho, s_\rho$ and $\eta_\rho$ are random variables, coupled
through Eq. (\ref{eq:timeaverage}) with each realization corresponding
to a realization of the single agent process
(\ref{eq:singleagent}). The variance of the zero-average Gaussian
variable $\eta_\rho$ can be obtained from (\ref{eq:correlcovariance})
and reads
\BE\label{eq:etasquared}
\bra \eta_\rho^2\ket_{\eta_\rho} &=&
\lim_{\tau,\tau^\prime\to\infty}\frac{1}{\tau\tau^\prime}\sum_{t\leq\tau}
\sum_{\tp\leq\tau^\prime}[(\id+G)^{-1}D(\id+G^T)^{-1}]_{t\tp}\nonumber\\
&=&\frac{\frac{\overline{\rho}}{1-\rhobar}+c}{(1+\chi)^2}.
\EE

The analysis now proceeds along the lines of \cite{HeimCool01}, and we
distinguish between so-called `frozen' and `fickle' agents. The
distinction between frozen and fickle agents was first introduced in
\cite{ChalMars99} and is based on observations from numerical
simulations of batch and on-line MGs. One finds that some of the
trajectories $q_i(t)$ of the original dynamics grow linearly in time
in the stationary state, $q_i(t)\sim t$, so that the corresponding
(frozen) agent always employs the same strategy. Other (fickle) agents
keep switching strategies, and their score difference remains
finite. While the distinction between frozen and fickle agents was
originally made on the level of the initial $N$-particle dynamics, the
same type of trajectories are also found in the realizations of the
effective single particle process (generated for example using the
method of
\cite{EissOppe92}). Frozen effective agents can be identified as those
realizations of (\ref{eq:singleagent}) for which $\tilde q_\rho\neq
0$. One then has $s_\rho=\sgn[\tilde q_\rho]$. Using
Eq. (\ref{eq:timeaverage}) (setting $\theta_\rho=0$) this is seen to
be the case if $|\eta_\rho|>\gamma_\rho$, where
$\gamma_\rho=\sqrt{\frac{\alpha(1-\rho)}{1-\overline{\rho}}}\frac{1}{1+\chi}$.
On the other hand, a given realization of the representative agent
process is `fickle' when the score-difference $q_\rho(t)$ fluctuates
around a finite value, with occasional zero-crossings, then one has
$\tilde q_\rho=0$. This is the case if $|\eta_\rho|<\gamma_\rho$, and
then $s_\rho=\eta_\rho/\gamma_\rho$. Note in particular that for
fickle (effective) agents $s_\rho$ is a continuous variable,
$-1<s_\rho<1$, with a non-zero value indicating that the corresponding
effective agent employs his or her strategies with different
frequencies.

From this the fraction of frozen agents with strategy correlation
$\rho$ can be computed\footnote{$\phi(\rho)$ is the probability for an
agent with strategy correlation $\rho$ to be frozen.} by performing
the Gaussian integral over $\eta_\rho$ (with variance given by
(\ref{eq:etasquared})):
\BE\label{eq:defphiofrho}
\phi(\rho)&=&\bra
\theta(|\eta_\rho|-\gamma_\rho)|\rho\ket_{\eta_\rho}\nonumber\\
&=&1-\erf\left(\frac{\lambda(\rho)}{\sqrt{2}}\right),
\EE
where we have introduced
\be\label{gl:lambdaofrho}
\lambda(\rho)=\sqrt{\frac{\alpha(1-\rho)}{\overline{\rho}+(1-\overline{\rho})c}}.
\ee
$\theta$ in (\ref{eq:defphiofrho}) is the step function; $\theta(x)=1$
if $x>0$, and $\theta(x)=0$ otherwise. The corresponding persistent
part of the correlation function reads
\BE
c(\rho)&\equiv&\lim_{t\to\infty}t^{-1}\sum_{\tau\leq t} C_\rho(\tau) \nonumber \\
&=&\bra s_\rho^2\ket_{\eta_\rho} \nonumber \\
&=& \bra \theta(|\eta_\rho|-\gamma)\ket_{\eta_\rho} +\bra (\eta_\rho/\gamma_\rho)^2
\theta(\gamma_\rho-|\eta_\rho|)\ket_{\eta_\rho} \nonumber \\
&=&
\phi(\rho)+\frac{1-\phi(\rho)}{\lambda(\rho)^2}-\frac{1}{\lambda(\rho)}
\sqrt{\frac{2}{\pi}}\exp\left(-\lambda(\rho)^2/2\right).\label{gl:cofrho}
\EE
Applying a static perturbation $\theta_\rho$ in (\ref{eq:timeaverage})
is (up to a prefactor $\sqrt{\alpha(1-\rho)/(1-\rhobar)}$) identical
to perturbing the static random variable $\eta_\rho$. Such
perturbations applied in the stationary state affect only fickle
agents, for which we have $s_\rho=\eta_\rho/\gamma_\rho$. The
susceptibility $\chi(\rho)\equiv\sum_\tau G_\rho(\tau)=\frac{\partial }{\partial
\theta_\rho}\bra s_\rho\ket_{\eta_\rho}$ may therefore be written as
\BE
\chi(\rho)&=&\sqrt{\frac{(1-\overline{\rho})}{\alpha(1-\rho)}}\bra \frac{\partial
s_\rho}{\partial\eta_\rho}\ket_{\eta_\rho}\nonumber \\
&=&\gamma_\rho^{-1}\sqrt{\frac{(1-\overline{\rho})}{\alpha(1-\rho)}} \left(1-\phi(\rho)\right).\label{gl:chiofrho}
\EE
From these equations the overall persistent correlation $c$ and susceptibility $\chi$ can be computed self-consistently as 
\BE
c&=&\int d\rho
P(\rho)\frac{1-\rho}{1-\rhobar}c(\rho), \label{eq:ctot} \\
\chi&=&\int d\rho P(\rho)\frac{1-\rho}{1-\rhobar}\chi(\rho)\label{eq:chitot}.
\EE
To this end, note that inserting
(\ref{eq:defphiofrho}) and (\ref{gl:lambdaofrho}) into
(\ref{gl:cofrho}) allows one to express $c(\rho)$ in terms of $c$ (and the
model parameters $\alpha$ and $\overline{\rho}$).

\section{Unimodal distribution of strategy correlations}

We will now first investigate the case of a unimodal distribution of
strategy correlations, $P(\rho^\prime)=\delta(\rho-\rho^\prime)$,
i.e. the case where all $\rho_i$ are equal, $\rho_i\equiv\rho$, with $0\leq\rho<1$.

\begin{figure}
\resizebox{0.4\textwidth}{!}{%
  \includegraphics{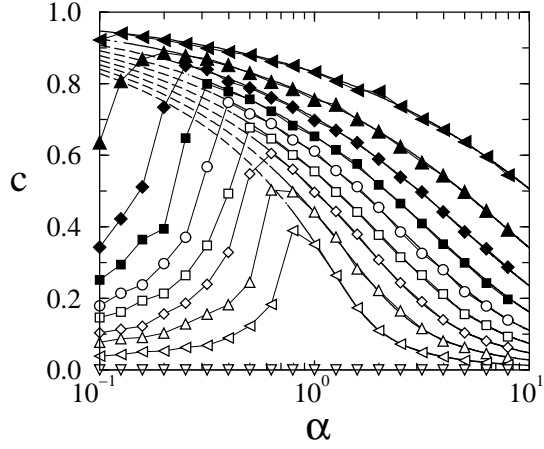}
}
\vspace{1em}
\caption{Persistent part $c$ of the correlation function for the
batch game with $\rho_i=\rho$ for all $i$ and with {\em tabula rasa}
initial conditions. Connected symbols are data obtained from simulations for
$N=300$ players, measured over $500$ batch steps, preceded by an
equilibration period of $500$ steps. The curves are
$\rho=0.9,0.8,\dots,0$ from top to bottom. The solid lines are the
theoretical predictions for the ergodic regime and have been continued
as dashed lines by extrapolation of equations (\ref{selfconsc},
\ref{selfconschi}) into the non-ergodic phase below $\alpha_c$, where
they are no longer valid. The slight discrepancy between the
theoretical lines and the numerical data close to the predicted
breakdown of the ergodic theory is due to finite-size effects.}
\label{fig:batch_c_phi}
\end{figure}

\subsection{Stationary state in the ergodic regime and phase diagram}

\begin{figure}
\resizebox{0.4\textwidth}{!}{%
  \includegraphics{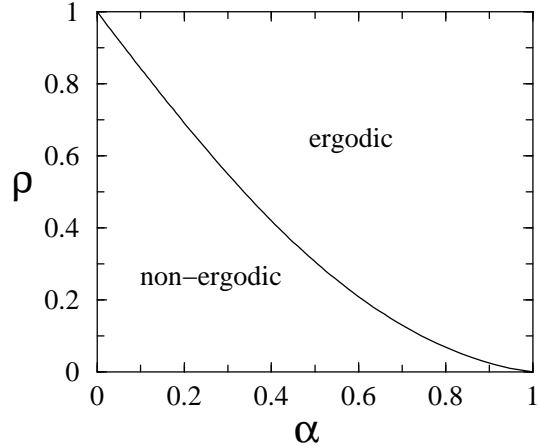}
}
\vspace{1em}
\caption[fig_phasediagram]{Phase diagram for the batch minority game
with unimodal distribution of strategy correlations,
$P(\rho^\prime)=\delta(\rho^\prime-\rho)$.\label{fig_phasediagram} }
\end{figure}

In this case the above self-consistent set of equations
(\ref{eq:ctot}, \ref{eq:chitot}) for the order parameters $c$ and
$\chi$ can be compactified considerably using equations
(\ref{gl:lambdaofrho}, \ref{gl:cofrho}, \ref{gl:chiofrho}) and we
find (with $c=c(\rho),\, \phi=\phi(\rho),\, \chi=\chi(\rho)$ and $\lambda=\lambda(\rho)$):
\begin{eqnarray}
c&=&1-\left(1-\frac{1}{\lambda^2}\right)\erf\left(\frac{\lambda}{\sqrt{2}}\right)\nonumber\\
&&-\frac{1}{\lambda}\sqrt{\frac{2}{\pi}}e^{-\lambda^2/2},\label{selfconsc}\\
\chi^{-1}&=&\frac{\alpha}{\erf\left(\frac{\lambda}{\sqrt{2}}\right)}-1\label{selfconschi},
\end{eqnarray}
where 
\begin{equation}
\lambda=\sqrt{\frac{\alpha}{\left(\frac{\rho}{1-\rho}+c\right)}}\label{selfconsy}.
\end{equation}
Note that $\rhobar=\rho$ in the case of unimodal
$P(\rho^\prime)=\delta(\rho-\rho^\prime)$. The fraction of frozen
agents is given by $\phi=1-\erf(\lambda/\sqrt{2})$.
Eqs. (\ref{selfconsc})-(\ref{selfconschi}) agree with the
corresponding equations for the static order parameters obtained
within a replica symmetric theory \cite{ChalMarsZhan00}. They are
easily solved numerically, and exact analytical predictions for $c,
\phi$ and $\chi$ as functions of $\alpha$ can be obtained for
$0\leq\rho<1$. Results for the persistent correlation $c$ are shown in
Fig. \ref{fig:batch_c_phi}, and we find very good agreement with
numerical simulations for large $\alpha$, greater than critical values
$\alpha_c(\rho)$. The deviations at lower values of
$\alpha=P/N<\alpha_c(\rho)$ are due to a breakdown of the ergodic
theory, more precisely of the assumption of finite integrated
response \cite{HeimCool01}. The point $\alpha_c(\rho)$ at which this happens can be
computed from Eq. (\ref{selfconschi}). One finds that
$\chi\to\infty$ at a critical value of $\alpha=\alpha_c(\rho)$ given by
$\alpha_c(\rho)=\erf(\lambda_c/\sqrt{2})$, where $\lambda_c=\lambda_c(\rho)$ fulfills

\begin{equation}
\erf\left(\frac{\lambda_c(\rho)}{\sqrt{2}}\right)=1+\frac{\rho}{1-\rho}-\sqrt{\frac{2}{\pi}}\frac{1}{\lambda_c(\rho)}e^{-\lambda_c(\rho)^2/2}.
\end{equation} 
Solving this equation numerically leads to the phase diagram depicted
in Fig. \ref{fig_phasediagram}, with the critical line in the
$(\alpha,\rho)$ plane separating the ergodic phase ($\chi$ finite) and
the non-ergodic phase. This phase diagram for the MG with diversified
strategies was first obtained from the corresponding replica
calculation in \cite{ChalMarsZhan00}.  While macroscopic observables
such as the volatility are sensitive to the initial conditions in the
non-ergodic phase, $\alpha<\alpha_c(\rho)$, the starting point is
irrelevant in the ergodic phase, $\alpha>\alpha_c(\rho)$. We display
the volatility for so-called {\em tabula rasa} starts ($q_i(0)=0$ for
all $i$), and for randomly biased starts ($|q_i(0)|=q_0={\cal O}(1)$)
in Fig. \ref{vol_batch_rho}.

\subsection{Analytic approximations for the volatility}

As we have seen it is possible to compute macroscopic order parameters
of the stationary state, such as the fraction of frozen agents or the
persistent part of the correlation function, exactly and in good
agreement with simulations in the ergodic regime for general values of
$\rho$. We will now turn to the magnitude of the market fluctuations
$\sigma^2$. By performing a direct average in the generating
functional, it was shown in \cite{HeimCool01} that the
disorder-averaged volatility matrix $\overline{\Xi}$ is proportional
to the correlator of the single-particle noise in the effective agent
process. Generalizing the results of
\cite{HeimCool01} to arbitrary values of $\rho$ we find the following
exact result:
\be\label{eq:exactvolbatch}
\overline{\Xi}_{t\tp}=(1-\rho)\Lambda_{t\tp}=(1-\rho)\left[(\id+G)^{-1}D(\id+G^T)^{-1}\right]_{t\tp}.
\ee
Thus, an exact calculation of the volatility requires the computation
of this matrix convolution, and hence the knowledge of both the
long-term and transient behaviours of the dynamical order parameters
$G(\tau)$ and $C(\tau)$ in the stationary state. To this end a full computation of
the time-translation invariant solutions of the self-consistent
representative agent problem would be necessary. Due to the retarded
self-interaction and the presence of the coloured single-particle
noise, this is in general impossible. One therefore has to resort to
approximations, and the generally accepted approach first presented in
\cite{HeimCool01} aims at expressing the market volatility in terms
of the persistent order parameters $c, \phi$ and $\chi$.

To this end one separates the contribution of the frozen agents to the
correlation function $C$ from the contribution of the fickle agents
and writes
\be
C(t-\tp)=\phi+(1-\phi)\bra s_\rho(t)s_\rho(\tp)\ket_{fi},
\ee 
where the average $\bra\dots\ket_{fi}$ extends only over fickle
agents. Inserting this into (\ref{eq:exactvolbatch}) and recalling\\
$\sigma^2=\lim_{T\to\infty}T^{-1}\sum_{t\leq T}\overline{\Xi}_{tt}$
leads to
\BE\label{eq:instantaneous}
\frac{1}{1-\rho}\sigma^2&=&\frac{\frac{\rho}{1-\rho}+\phi}{(1+\chi)^2}
+(1-\phi)\nonumber\\&&\times\lim_{\tau\to\infty}\tau^{-1}\sum_{t\leq\tau}\bra\left[\sum_{\tp\leq
t}(\id+G)^{-1}_{t\tp}s_\rho(\tp)\right]^2\ket_{fi}.
\EE
In \cite{HeimCool01}, for $\rho=1/2$, the authors proceeded by keeping only the
instantaneous contribution, $t=t^\prime$, in the last
term. Generalizing their calculation to arbitrary $0\leq\rho<1$ one
obtains the following approximate expression
\begin{equation}\label{eq:vol_analytical}
\sigma^2=(1-\rho)\left[\frac{\rho/(1-\rho)+\phi}{(1+\chi)^2}+(1-\phi)\right].
\end{equation}

Fig. \ref{vol_batch_rho} demonstrates that this approximation is in
good agreement with data from simulations of the original batch
process for $\rho\gtrsim 1/2$, but that it becomes less accurate as
$\rho$ is reduced. In particular we note that in simulations we find
volatilities $\sigma^2>1$ in the ergodic phase of batch games with
largely anti-correlated strategies, i.e. low values of $\rho$, while
(\ref{eq:vol_analytical}) predicts a volatility below the random
trading limit $\sigma^2=1$ for all values of $\rho$ and
$\alpha\geq\alpha_c(\rho)$. We have verified explicitly that
the discrepancy between this analytical approximation and the
numerically measured volatility can be traced back to the omission of
the non-instantaneous terms in (\ref{eq:instantaneous}): iterating the
single-particle process using the Eissfeller-Opper algorithm
\cite{EissOppe92}, measuring the contributions from the frozen and
fickle agents, respectively, and taking into account all terms of
(\ref{eq:instantaneous}) restores the agreement with direct numerical
simulations of the batch process. The discrepancy between the
approximation of Eq. (\ref{eq:vol_analytical}) and the volatility
measured in simulations becomes extremal in the fully anti-correlated
case. In this case one finds no frozen agents, and inserting $\rho=0$
and $\phi=0$ into (\ref{eq:vol_analytical}) gives $\sigma^2=1$ for
$\alpha>\alpha_c(\rho=0)=1$. In numerical simulations of the batch
process, however, we find a decreasing relation
$\sigma^2=\sigma^2(\alpha)$, and $\sigma^2=1$ is only approached in
the limit $\alpha\to\infty$, see Fig. \ref{fig:biasedbatchrho0}. In
the next section, we will study the case of fully anti-correlated
strategies in more detail, and will derive different analytical
approximations for the global market fluctuations in this case.

\subsection{The fully anti-correlated case}

\begin{figure}
\resizebox{0.38\textwidth}{!}{%
  \hspace{1cm}\includegraphics{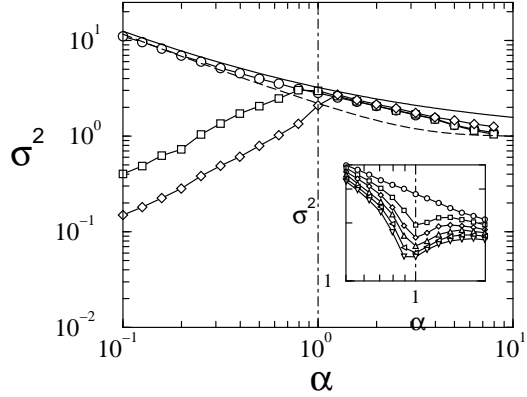} }
\vspace{1em}
\caption{Volatility as a
function of $\alpha$ for $\rho=0$ for the batch game. Connected
symbols are from simulations started from different initial conditions
(circles are {\em tabula rasa} starts, $|q_i(0)|=0$, squares are
$|q_i(0)|=2.0$ and diamonds $|q_i(0)|=10.0$). The solid and dashed lines
are the approximations of Eqs. (\ref{eq:volapproxrho0}) and
(\ref{eq:volapproxrho0_2}) respectively. The vertical dot-dashed line
marks the analytically predicted location of the phase transition at
$\alpha_c(\rho=0)=1$. Inset: $\sigma^2$ vs $\alpha$ for $\rho=0.0,
0.01, 0.02, 0.03, 0.04, 0.05$ from top to bottom ({\em tabula rasa}
starts).
\label{fig:biasedbatchrho0} }
\end{figure}
In the special case of completely anti-correlated strategies,
$\rho=0$, and {\em tabula rasa} starts, we observe experimentally that
the system is always in a fully oscillating phase, i.e. that all
agents switch strategies at each time step, and that accordingly
$C(\tau)=(-1)^\tau$ for all $\tau$. Eq. (\ref{selfconsc}) for $\rho=0$
is indeed solved by $c=0$ so that no frozen agents are possible. We
will use this observation as an ansatz, which will allow us to proceed
analytically and to give an approximate expression for the volatility,
which is different from the one (derived for general $\rho$) in the
previous section. Its derivation is similar to the analysis of the
non-ergodic state in the limit $\alpha\to 0$ of the standard batch MG
\cite{HeimCool01}. In this limit of the game with uncorrelated strategies a similar oscillatory state is
found.

For $\rho=0$, we have $D(\tau)=C(\tau)=(-1)^\tau$ and using
Eq. (\ref{eq:exactvolbatch}), we conclude that the correlation matrix
of the single-particle noise in the stationary state is given by
$\Lambda_{t\tp}=(-1)^{t-\tp}\Upsilon^2,$ where $\Upsilon=\sum_\tau
(-1)^\tau (\id+G)^{-1}(\tau)$. Thus, for any fixed realization of the
single-particle noise $\eta_0(t)$ must be of the form
$\eta_0(t)=\Upsilon z (-1)^t$, where $z$ is a (static) Gaussian random
variable of zero mean and unit variance (note that the subscript in
$\eta_0(t)$ indicates that we are concerned with the case $\rho=0$ in
this section, similarly for $q_0(t)$ and $\tilde s_0$ below). A given
stochastic value of $z$ determines the oscillation amplitude and sign
of the corresponding trajectory of the noise $\eta_0(t)$. The
volatility in this fully oscillatory state can be obtained as
$\sigma^2=\lim_{T\to\infty}T^{-1}\sum_{t\leq
T}\Lambda_{tt}=\Upsilon^2$, so that it remains to compute
$\Upsilon$. Upon introduction of
\be
\tilde s_0 \equiv \lim_{\tau\to\infty}\tau^{-1}\sum_{t\leq\tau} (-1)^t \sgn[q_0(t)]
\ee
we can use the identity
\be
\sqrt{\alpha}\sum_{\td}\bra\eta_0(t)\eta_0(\td)\ket_{\eta_0}
G_{t^\prime
\td}=\bra\sgn[q_0(t^\prime)]\eta_0(t)\ket_{\eta_0}
\ee (obtained by an integration by
parts in the generating functional \cite{HeimCool01}) to write $\Upsilon$ as
\begin{equation}
\Upsilon=1-\frac{1}{\sqrt{\alpha}}\bra\tilde s_0 z\ket_z,
\end{equation}
where $\bra\dots\ket_z$ denotes an average over the standard Gaussian variable $z$. We have also used the fact that $\sum_\tau (-1)^\tau(\id+G)^{-1}(\tau)=[1+\sum_\tau(-1)^\tau G(\tau)]^{-1}$ here \cite{HeimCool01}.

In order to compute the average $\bra \tilde s_0 z\ket_z$, we note that from the
effective single-agent process we can derive the relation
\begin{equation}\label{eq:eqforoscagent}
\widetilde{\Delta q_0}=-\alpha\Upsilon\tilde s_0+\sqrt{\alpha}\tilde\eta_0
\end{equation}
between the staggered averages $\tilde s_0$ and
\begin{eqnarray}
\widetilde{\Delta q_0}&=&\lim_{\tau\to\infty}\tau^{-1}\sum_{t\leq\tau} (-1)^t(q_0(t+1)-q_0(t)),\\
\tilde \eta_0 &=&
 \lim_{\tau\to\infty}\tau^{-1}\sum_{t\leq\tau} (-1)^t \eta_0(t).
\end{eqnarray}
Since the effective agents are purely oscillatory, one has
$\sgn(\widetilde{\Delta q_0})=-\tilde s_0$. Using
Eq. (\ref{eq:eqforoscagent}) and the relation $\eta_0(t)=\Upsilon
z(-1)^t$ we therefore require $\sgn[z]=-\tilde s_0$, i.e. $\tilde s_0
z=-|z|$ for any given realization of $z$ with
$\sqrt{\alpha}\Upsilon|z|>\alpha\Upsilon$ (and the corresponding
realization of the effective agent process). Since $\Upsilon$ comes
out positive this is the case for realizations with
$|z|>\sqrt{\alpha}$. For $|z|<\sqrt{\alpha}$ both signs $\tilde s_0
z=\pm |z|$ are possible; using the Eissfeller-Opper algorithm we have
checked that indeed both signs are taken in this case. A
similar complication has been encountered in the context of mixed
Minority/Majority Games in \cite{DeMaGiarMose03}.
\begin{figure}[t]
\resizebox{0.4\textwidth}{!}{%
  \includegraphics{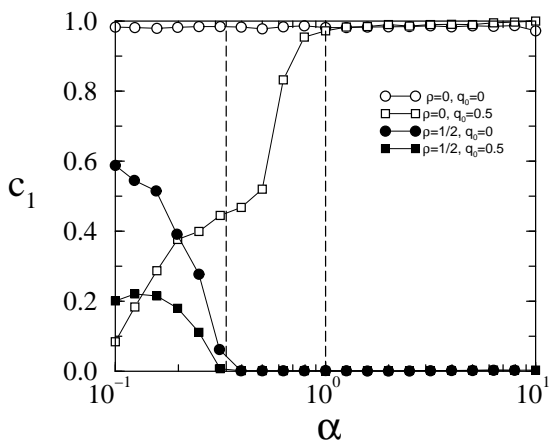} }
\vspace{1em}
\caption{Amplitude of oscillations of the correlation function as a function of $\alpha$ for the batch game for $\rho=1/2$ (filled markers) and $\rho=0$ (open markers). Results are from simulations with $N=500$ agents, run for $500$ batch steps and averaged over $10$ realisations of the disorder. In the simulations we measure $c_1$ in the stationary state as $c_1=(2T)^{-1}\sum_{\tau=T}^{2T} |C(\tau+1)-C(\tau)|$, where $T=125$ batch steps. Circles are data for {\em tabula rasa} starts, squares from simulations with biased starts ($q_0=0.5$). The vertical dashed lines mark the locations of the phase transition of the games with uncorrelated strategies at $\alpha_c(\rho=1/2)\approx 0.3374$ and at $\alpha_c(\rho=0)=1$ for fully anticorrelated strategies.
\label{fig:oscillations} }
\end{figure}

We have two possibilities to proceed: firstly, we can assume that
$\tilde s_0 z=-|z|$ also for $-\sqrt{\alpha}<z<\sqrt{\alpha}$, and
that deviations from this behaviour will only have a small effect on
the volatility. We then end up with the following approximation
\begin{eqnarray}\label{eq:volapproxrho0}
\sigma^2&\approx& \left(1+\frac{1}{\sqrt{\alpha}}\bra |z|\ket_z\right)^2
\nonumber \\
&=& \left(1+\sqrt{\frac{2}{\pi\alpha}}\right)^2.
\end{eqnarray}
Alternatively, we can assume that both signs $\tilde s_0 z=\pm|z|$ are taken equally often for $|z|<\sqrt{\alpha}$ so that this interval does not contribute to the average $\bra \tilde s_o z\ket_z$.  We then find
\begin{eqnarray}\label{eq:volapproxrho0_2}
\sigma^2&\approx& \left(1+\sqrt{\frac{2}{\pi\alpha}}e^{-\alpha/2}\right)^2.
\end{eqnarray}
As depicted in the main panel of Fig. \ref{fig:biasedbatchrho0} the
approximations of (\ref{eq:volapproxrho0}) and
(\ref{eq:volapproxrho0_2}) appear to form upper and lower bounds of
the volatility, respectively. Both approximations become exact only in
the limit $\alpha\to 0$, but reproduce the behaviour of
$\sigma^2=\sigma^2(\alpha)$ quite well, even for larger values of
$\alpha\approx 1$.

We conclude this section by mentioning that the batch MG with fully
anti-correlated strategies, $\rho=0$, is different from the cases
$0<\rho<1$ in two respects: (i) the fully anti-correlated case
exhibits oscillatory behaviour, $C(\tau)=(-1)^\tau$, for all $\alpha$,
even above $\alpha_c(\rho=0)=1$, whereas in all other cases
oscillations decay above $\alpha_c$ and persist only in the
non-ergodic phase, where one finds correlation functions of the form
$C(\tau)=c_0+c_1(-1)^\tau$ for $\tau>0$\footnote{Below $\alpha_c$ we
find $C(\tau)=c_0+c_1(-1)^\tau$ for $\tau>0$ within the experimental
accuracy and with positive fitting parameters $c_0$ and $c_1$. We
observe that $c_0+c_1<1$ unless the case $\rho=0$ or the limit
$\alpha\to 0$ is considered. Trivially one always has
$C(\tau=0)=1$.}. This is illustrated in Fig. \ref{fig:oscillations},
where we plot $c_1$ as a function of $\alpha$ for different values of
$\rho$; (ii) the volatility plotted as a function of $\alpha$ exhibits
a minimum at $\alpha_c(\rho)$ for all $0<\rho<1$, but is decreasing
monotonically for $\rho=0$, see the inset of
Fig. \ref{fig:biasedbatchrho0}. Nevertheless the stationary volatility
in the fully anti-correlated case does depend on initial conditions
below $\alpha_c(\rho=0)=1$, whereas the starting point is irrelevant
for $\alpha>1$.

\subsection{Comparison with the on-line game}

\begin{figure}
\resizebox{0.4\textwidth}{!}{%
  \includegraphics{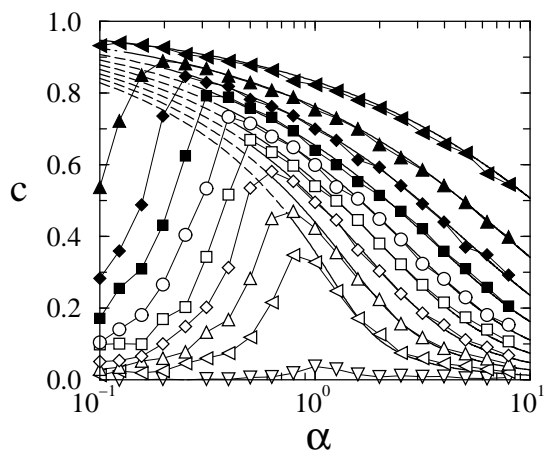}
}
\vspace{1em}
\caption{Persistent part $c$ of the correlation function for the
on-line game with $\rho_i=\rho$ for all $i$ and with {\em tabula rasa}
initial conditions. Connected symbols are data obtained from simulations for
$N=300$ players, measured over $25000$ on-line steps, preceded by an
equilibration period of $25000$ steps. The curves are
$\rho=0.9,0.8,\dots,0$ from top to bottom. The solid lines are the
theoretical predictions for the ergodic regime and have been continued
as dashed lines by extrapolation into the non-ergodic phase below
$\alpha_c$, where they are no longer valid. The slight discrepancy
between the theoretical lines and the numerical data close to the
predicted breakdown of the ergodic theory is due to finite-size
effects.}
\label{fig:c_online}
\end{figure}

\begin{figure}[t]
\resizebox{0.5\textwidth}{!}{%
  \includegraphics{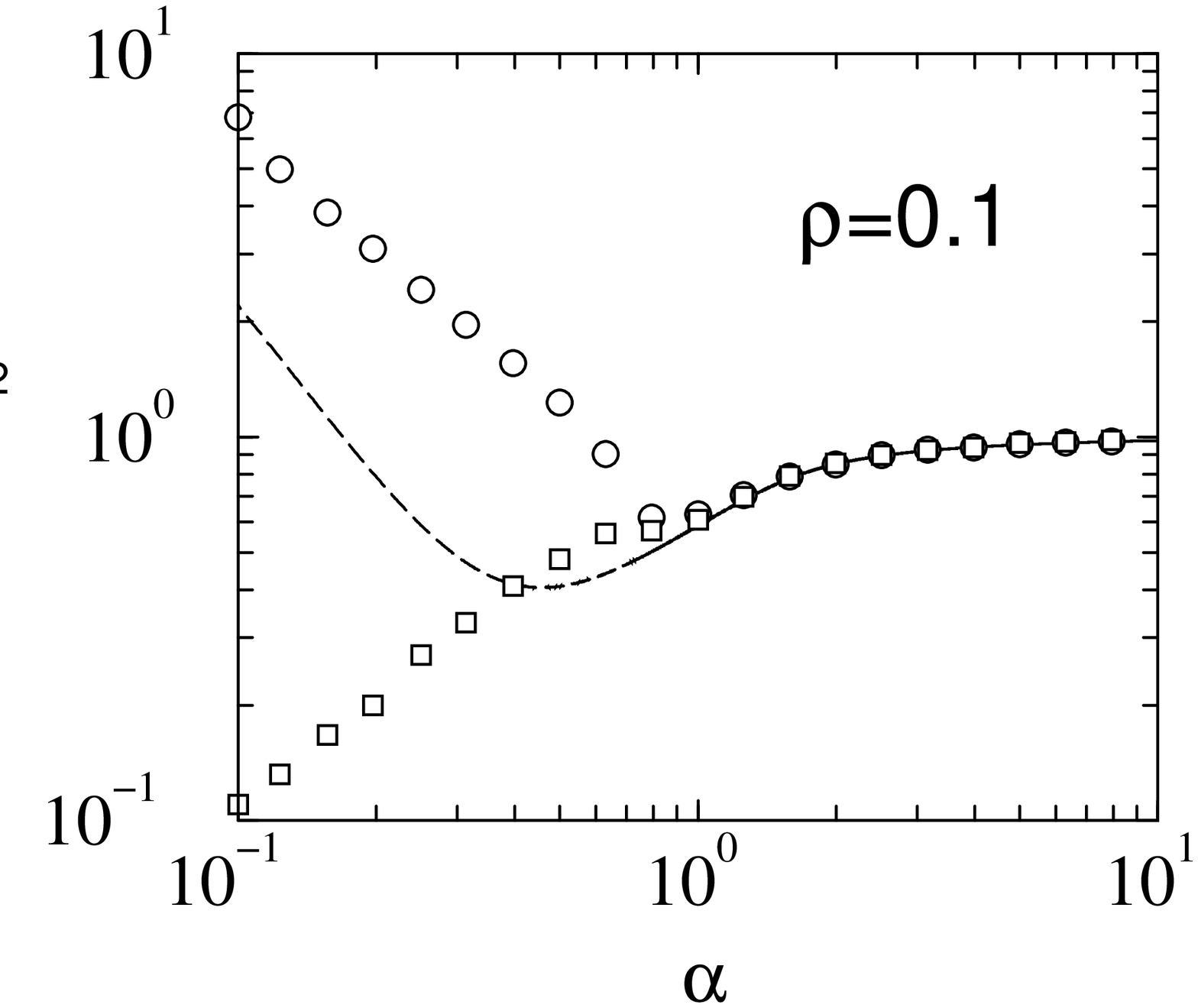} ~~~~
  \includegraphics{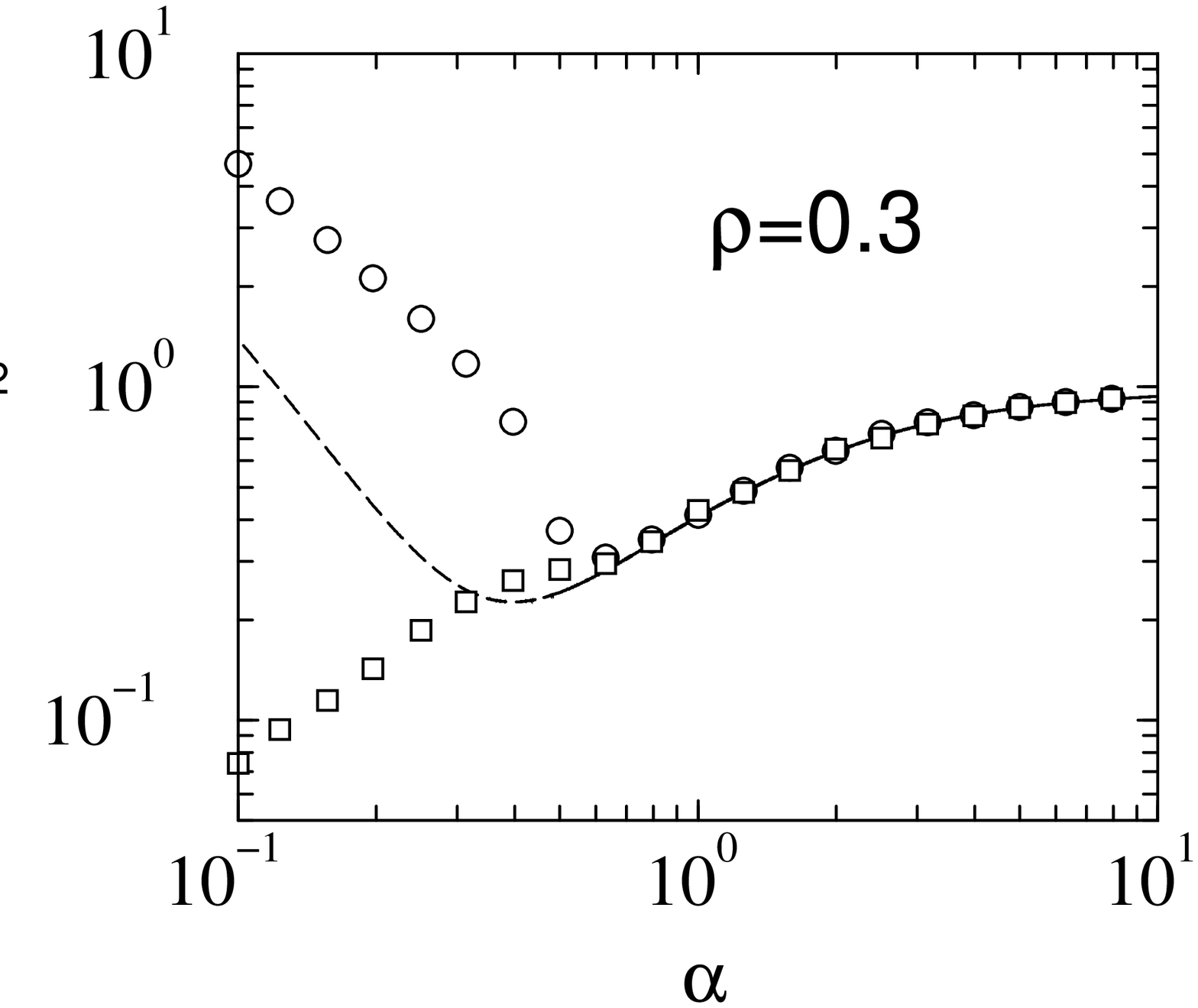}}\\
\resizebox{0.5\textwidth}{!}{%
 \includegraphics{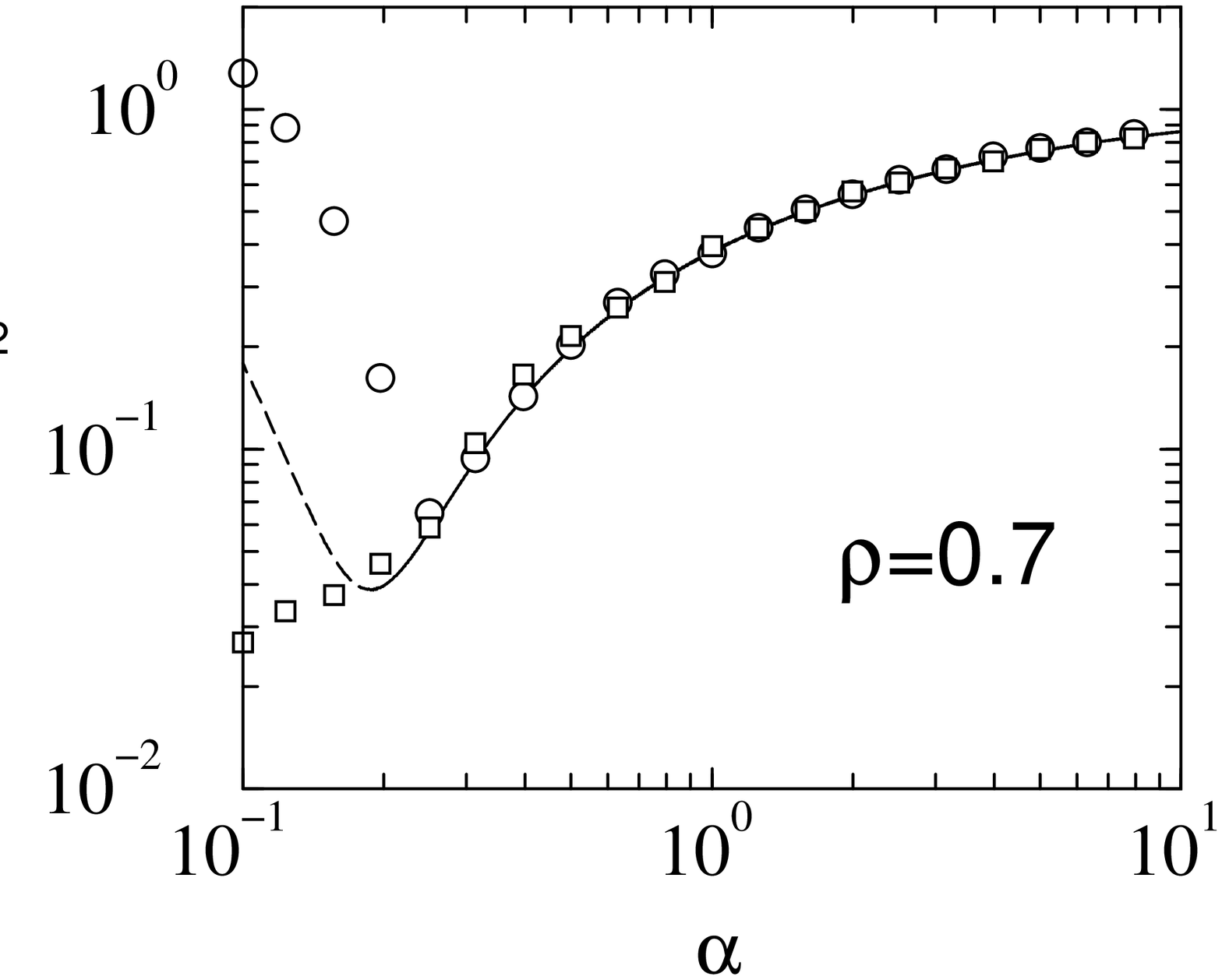} ~~~~
  \includegraphics{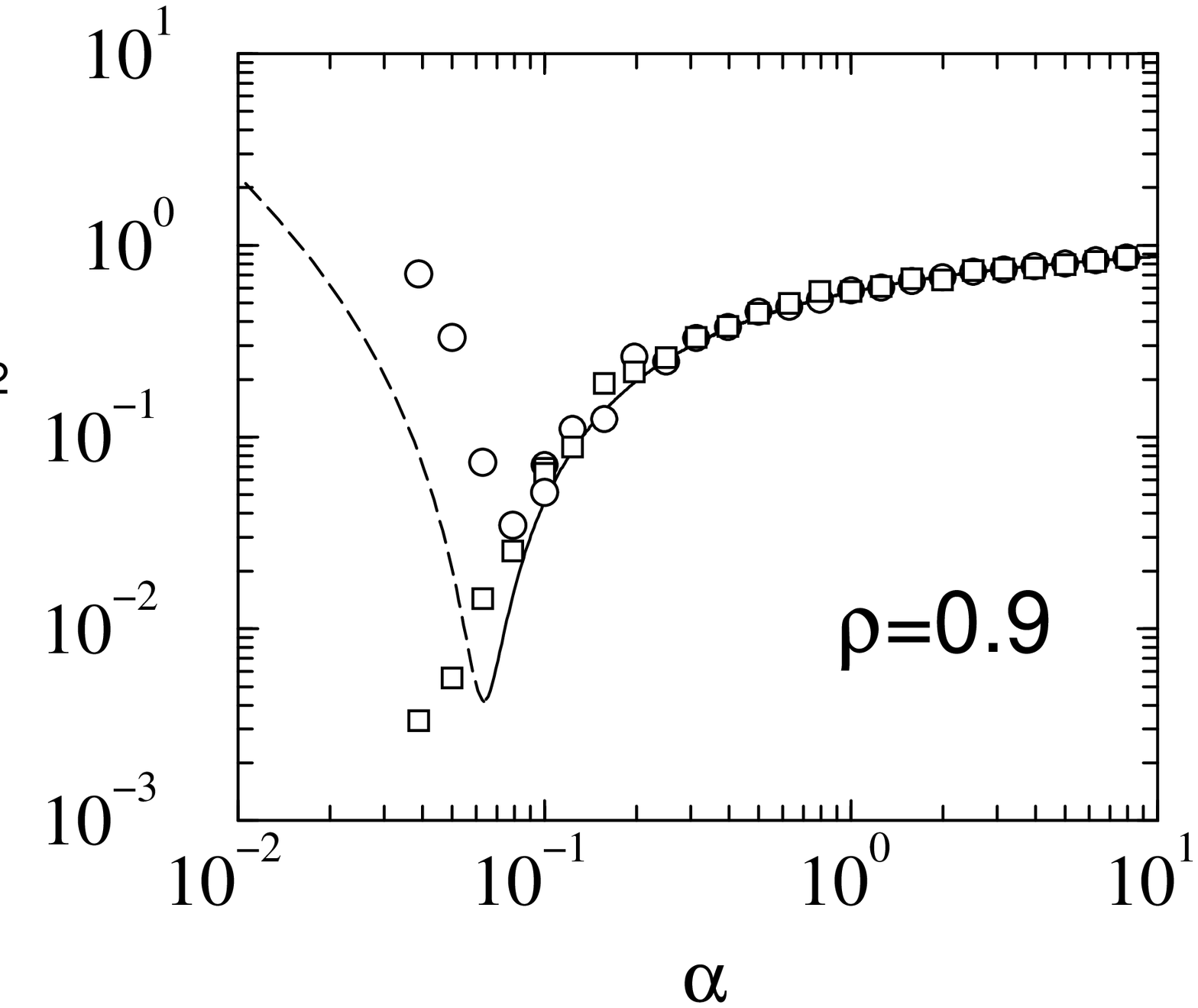}
}
\vspace{1em}
\caption{\label{fig:vol_online_panel}Volatility as a function
of $\alpha$ for different values of $\rho$ for the on-line game
($\rho_i=\rho$ for all agents). Solid
lines are the approximation of (\ref{eq:vol_analytical2}) for the
ergodic phase and are continued as dashed lines into the non-ergodic
phase (where the ergodic theory is no longer valid). Markers are from
simulations of the on-line process with $N=300$ agents ($N=1000$ for
$\rho=0.9$), averaged over $10$ samples of the disorder, and run for
$10^5$ on-line time steps. Circles: {\em tabula rasa start}
($q_i(0)=0$), squares: biased starts ($|q_i(0)|=3$).}
\end{figure}

We will now turn to the case of on-line learning rules, and will
discuss the influence of the timing of adaptation of the agents on the
behaviour of MGs with correlated strategies.

A generating functional theory for the on-line game with random
external information and uncorrelated strategies has been worked out
in \cite{CoolHeim01}. The analysis has to deal with the explicit
dependence on the external information, and inevitably requires a much
more elaborate mathematical formalism than the analysis of the batch
game. The resulting theory is easily adapted to the case of general
strategy assignments with arbitrary correlations. We will not report
the mathematical details of the calculation, but will only state
that, in the ergodic regime, it leads to self-consistent equations for
the persistent order parameters (such as $c,
\phi$ and $\chi$), which are identical to the ones of the batch MG with general strategy correlations. As a consequence, the location of the phase transition (marked by an onset of diverging integrated response) is identical in batch and on-line games with the same distribution $P(\rho)$ of correlation parameters. Results for the persistent correlation $c$ as a function
of $\alpha$ are compared with simulations of the on-line game with
unimodular strategy correlations in Fig. \ref{fig:c_online}. The on-line simulations shown in  Fig. \ref{fig:c_online} essentially also match those of the batch case shown in Fig. \ref{fig:batch_c_phi} for the ergodic region.

\medskip 

\begin{figure}[t]
\resizebox{0.4\textwidth}{!}{%
  \hspace{2.5cm}\includegraphics{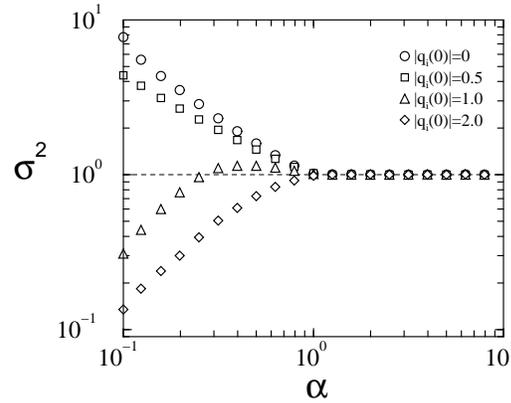}
}
\vspace{1em}
\caption{Volatility as a
function of $\alpha$ for the on-line game at $\rho=0$, and for
different values of the initial bias $|q_i(0)|$. All data is from simulations
with $N=300$ agents averaged over $10$ samples of the disorder.
\label{fig:onlinebiasedrho00} }
\end{figure}

\begin{figure}
\resizebox{0.4\textwidth}{!}{%
  \includegraphics{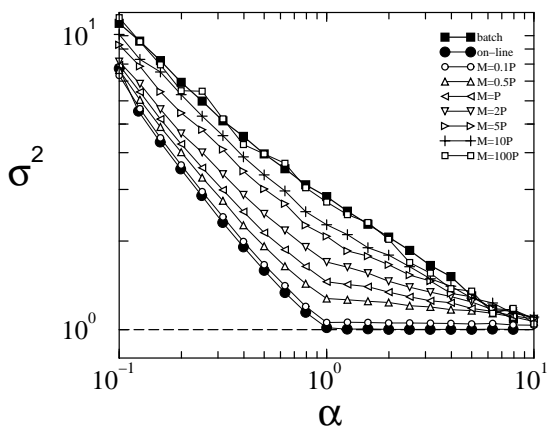}
}
\vspace{1em}
\caption{Volatility of a MG with fully anti-correlated strategies ($\rho_i=0$ for all $i$) and in which switching of strategies is allowed only every $M$ time steps. The individual curves are for different values of $M$. All data is from simulations performed at a constant number $P=\alpha N=50$ of possible states of the external information $\mu$ and were run for $10^6$ on-line steps. Results are averages over $20$ realisations of the disorder. The corresponding curves for the on-line and batch games are displayed as filled symbols for comparison. The lines connecting the markers are guides to the eye.}
\label{fig:online_delta_scan}
\end{figure}

Numerical simulations, however, reveal crucial differences between the
volatilities of the on-line and batch games with correlated
strategies.  A comparison of Figs. \ref{vol_batch_rho} and
\ref{fig:vol_online_panel} shows that in their ergodic states the volatility of batch games can take values both above and below the random trading
limit $\sigma^2=1$ , whereas that in on-line games never exceeds
$\sigma^2=1$. The difference between the volatilites of batch and
on-line games becomes maximally pronounced in the fully
anti-correlated case, compare Figs. \ref{fig:biasedbatchrho0} and
\ref{fig:onlinebiasedrho00}. In the on-line game with $\rho=0$ we find
that the volatility is equal to one for all
$\alpha>\alpha_c(\rho=0)=1$, whereas it approaches this random
trading limit only asymptotically for $\alpha\to\infty$ in the batch
case. Allowing the agents to switch stategies only every $M$ time
steps according to the update rule (\ref{eq:updatem}) one can
interpolate between the batch and on-line cases\footnote{Note also
that some work on mixed populations of agents with individual updating
frequencies $M_i$ is currently being finalised by other authors
\cite{Mosetti}.}: Fig. \ref{fig:online_delta_scan} shows the volatility
$\sigma^2$ as a function of $\alpha$ for the fully anti-correlated
case for different intermediate choices of $M$. While small values of
$M\ll P$ produce market fluctuations close to those of the on-line
game (which corresponds to $M=1$), the batch limit is obtained for
$M\geq{\cal O}(P)$.

Similarly to the batch case, a computation of the transient parts of the
dynamical order parameters would be necessary to obtain exact
expressions for the volatility of on-line MGs. As this is in general
not feasible, one proceeds as in the batch case, and tries to find
approximate expressions in terms of the persistent order parameters
only. Such an approach was first proposed for on-line games in
\cite{CoolHeim01}, and is slightly different from the approximations
in the batch case. The approximation is based on the observation that
in mean-field disordered systems with detailed balance
fluctuation-dissipation relations can be used to `transform away'
non-persistent parts of the response and correlation functions without
changing averages of single-time observables and averages of two-time
quantities at infinite time-separations. For equilibrium systems with
detailed balance this approach is exact \cite{Cool00b}. It can be used
to extract the static order parameters as obtained in a
replica-symmetric equilibrium approach from the generating functional
equations. Details of this procedure can be found in
\cite{Cool00b,HeimelThesis}.

Although the MG does not exhibit detailed balance a similar {\em
ad-hoc} procedure for removing the non-persistent parts of the order
parameters has been successfully employed to obtain approximations for
the volatility in \cite{CoolHeim01}. It amounts to assuming that $C$
and $G$ are of the following form:
\be
C_{t\tp}=c+(1-c)\delta_{t\tp},~~~~~~~~~~~
G_{t\tp}=\chi\delta(1-\delta)^{t-\tp-1} \, (\mbox{for }t>\tp),
\ee
with the limit $\delta\to 0$ to be taken at the end
of the calculation. In this way one preserves\\
$\lim_{\tau\to\infty}C(\tau)=c$ and $\sum_{\tau}G(\tau)=\chi$, but removes
non-persistent contributions.
 
This approach is easily generalised to the case of unimodular
distributions of the correlation parameter, and upon making the standard
assumptions on time-translation invariance and ergodicity, one obtains
the following approximate expression for the volatility
\begin{equation}\label{eq:vol_analytical2}
\sigma^2=(1-\rho)\left[\frac{\rho/(1-\rho)+c}{(1+\chi)^2}+(1-c)\right],
\end{equation}
which differs only slightly from the approximate result for the
batch game, Eq.  (\ref{eq:vol_analytical}). As shown in
Fig. \ref{fig:vol_online_panel} this approximation is accurate in the
ergodic phase for all values of $\rho$, even in a regime where the
corresponding approximation for the batch MG is not satisfactory to
describe the volatility measured in numerical experiments. We
attribute the deviations slightly above the transition to finite-size
effects. Applying the same method to the volatility of the batch game also leads to Eq.  (\ref{eq:vol_analytical2}) and hence to the same qualitative and quantative discrepancy between the numerically measured and analytically estimated volatilities in batch games with largely anti-correlated strategies. See also \cite{MGbook2} for alternative derivations of Eq. (\ref{eq:vol_analytical2}) for batch and on-line games with uncorrelated strategies.

\medskip

Eq. (\ref{eq:vol_analytical2}) agrees with the approximation for the
volatility derived from the replica analysis of the MG with
diversified strategies \cite{ChalMarsZhan00}. We note that the
analogue of the removal of the non-persistent parts of $C$ and $G$ in
the dynamical approach is reflected by an assumption on the
cross-correlations between the agents in the replica formalism of
\cite{ChalMarsZhan00}. Based on the assumption that $\bra
s_is_j\ket=\bra s_i\ket\bra s_j\ket$ for $i\neq j$ the authors of \cite{ChalMarsZhan00}
neglect a term $\Delta\equiv N^{-1}\sum_{i\neq j}\sum_\mu
\xi_i^\mu\xi_j^\mu(\bra s_i s_j\ket-\bra s_i\ket\bra s_j\ket)$, in the ergodic
phase ($\alpha>\alpha_c$). Here again $s_i(t)=\sgn[q_i(t)]$, and
$\bra\dots\ket$ denotes an average over time. While our findings
concerning the volatility suggest that this assumption is appropriate
in the on-line game, it appears to be inaccurate in the batch game at
low values of $\rho$. We have confirmed this numerically in
simulations of the batch MG. At fixed $\alpha>1$ we find that
$\Delta=\Delta(\alpha,\rho)$ is close to zero for large values of
$\rho\simeq 1$, but increases as $\rho$ approaches zero. While we do
not depict these results here, we will only point out that for
$\rho=0$ we find oscillatory behaviour, $s_i(t)=s_i(0)(-1)^t$, at all
$\alpha$, so that $\bra s_i\ket=0$, but $\bra s_i s_j\ket=\pm 1$. One then has
$\bra s_is_j\ket-\bra s_i\ket \bra s_j\ket=\pm 1 \neq 0$ so that the above approximation
fails.

To conclude the discussion of the volatility in on-line models, we
would like to mention that the volatility of MGs with real histories
and fully anti-correlated strategies behaves qualitatively like the
one of the on-line MG with random history, with $\sigma^2=1$ for $\alpha>1$
\cite{ChalletThesis}.

\medskip

Finally, in this section, let us briefly address the role of global
oscillations in on-line MGs. As an analytical treatment of this
dynamical feature would require a solution of the transient behaviour
of the dynamical order parameters (which is still awaited) the results
in the remainder of this section and in section \ref{sec:randomtiming}
are all based on direct numerical simulations of the games under
consideration. Anti-persistent behaviour in the on-line MG with
uncorrelated strategies and with real market histories was identified
first in
\cite{ChalMars99}, where the authors find oscillatory behaviour below $\alpha_c$, but no persistent oscillations in the high-$\alpha$ phase, in qualitative agreement with results for batch games \cite{HeimCool01}, recall also Fig. \ref{fig:oscillations}. In \cite{ChalMars99} the authors considered an on-line
version of the MG in which the agents use only the sign of the total
bid to update their strategy scores, i.e. a model, in which
$A(\ell^\prime)$ in Eq. (\ref{update}) is replaced by
$\pi(\ell^\prime)=\sgn[A(\ell^\prime)]$. We will refer to this as the
`sign-update' rule, as opposed to the linear relation
(\ref{update}). The authors then consider a conditional correlation
function
\be
C_{\pi\pi}(\tau)=P^{-1}\sum_{\mu=1}^P \bra \pi(\ell)\pi(\ell+\tau)|\mu\ket
\ee
in the stationary state, where the average $\bra \pi(\ell)\pi(\ell+\tau)|\mu\ket$ extends only over times $\ell$ and $\ell+\tau$ for which $\mu(\ell)=\mu(\ell+\tau)=\mu$. Similarly we can define a conditional `spin-spin' correlation function according to
\be
C_{ss}(\tau)=(NP)^{-1}\sum_{\mu=1}^P\sum_{i=1}^N \bra s_i(\ell)s_i(\ell+\tau)|\mu\ket.
\ee
\begin{figure}
\resizebox{0.5\textwidth}{!}{%
  \includegraphics{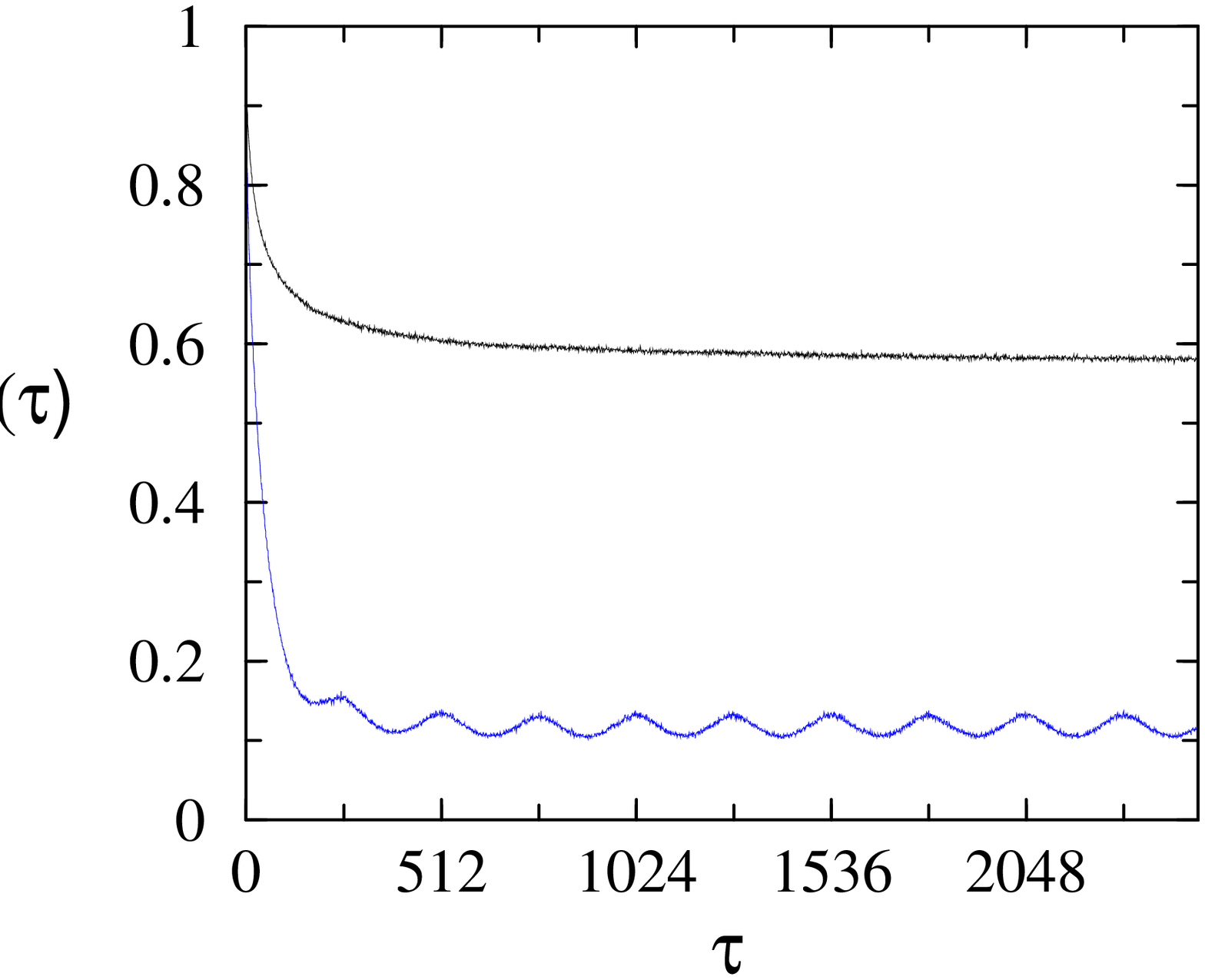} ~~~~
  \includegraphics{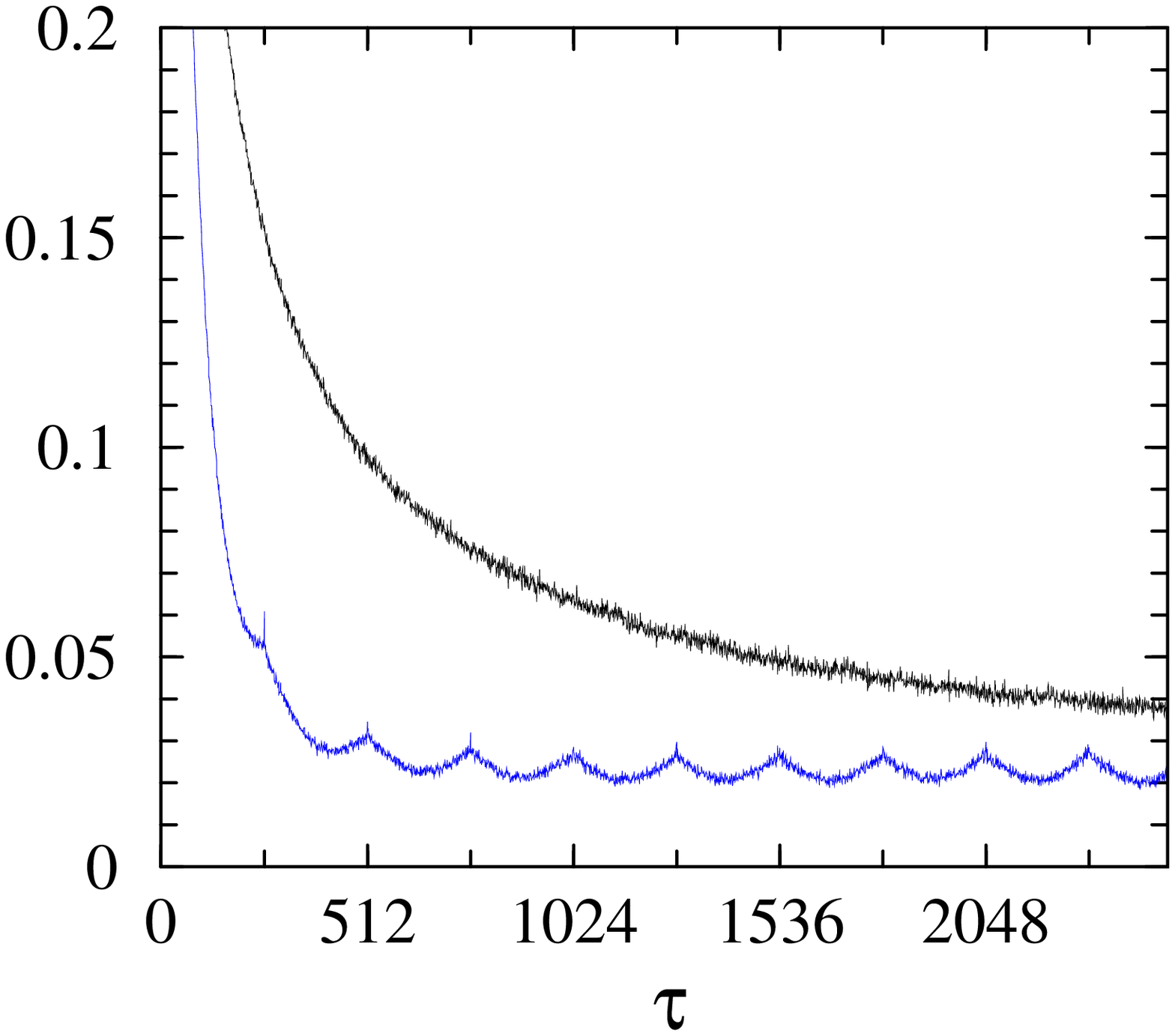}}\\
\resizebox{0.5\textwidth}{!}{%
 \includegraphics{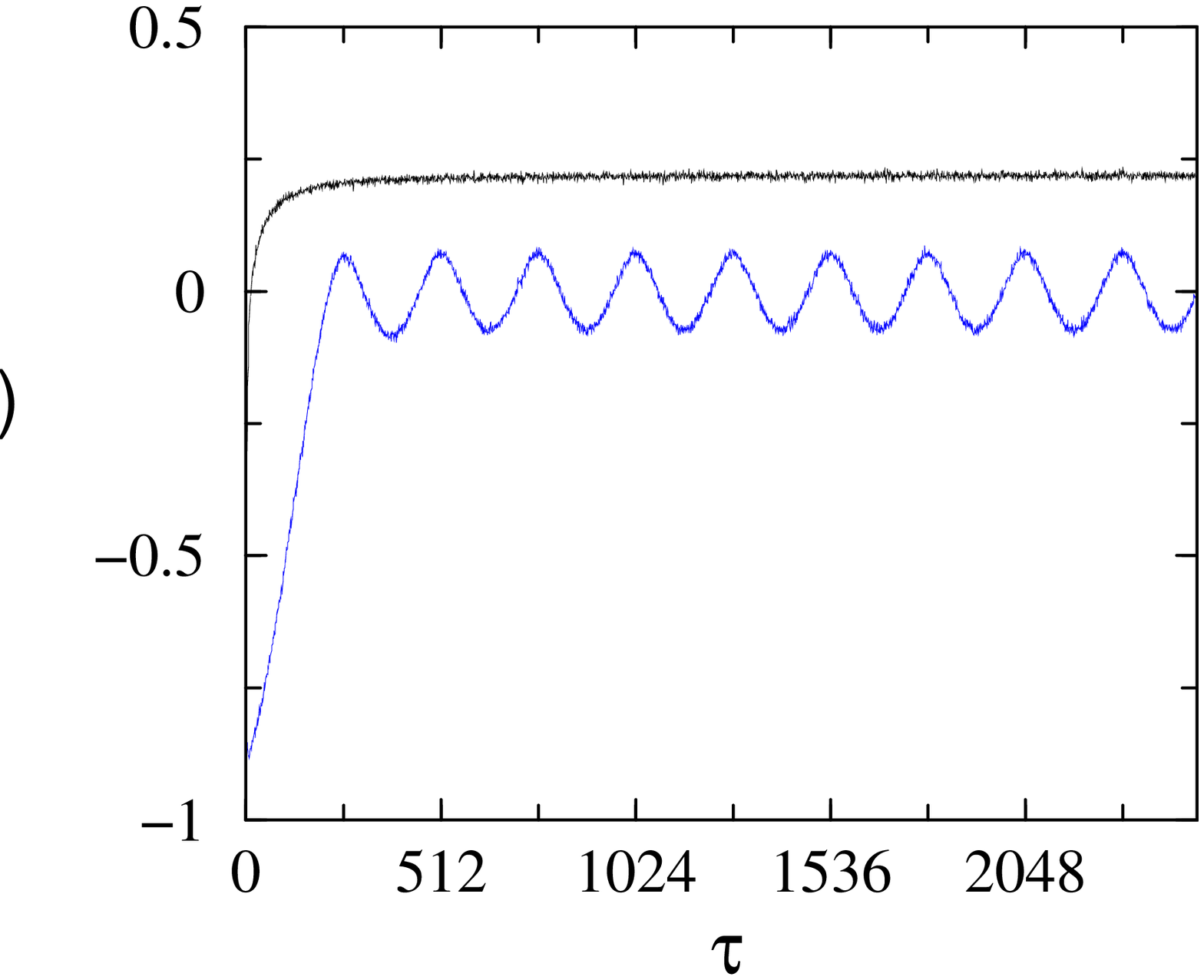} ~~~~
  \includegraphics{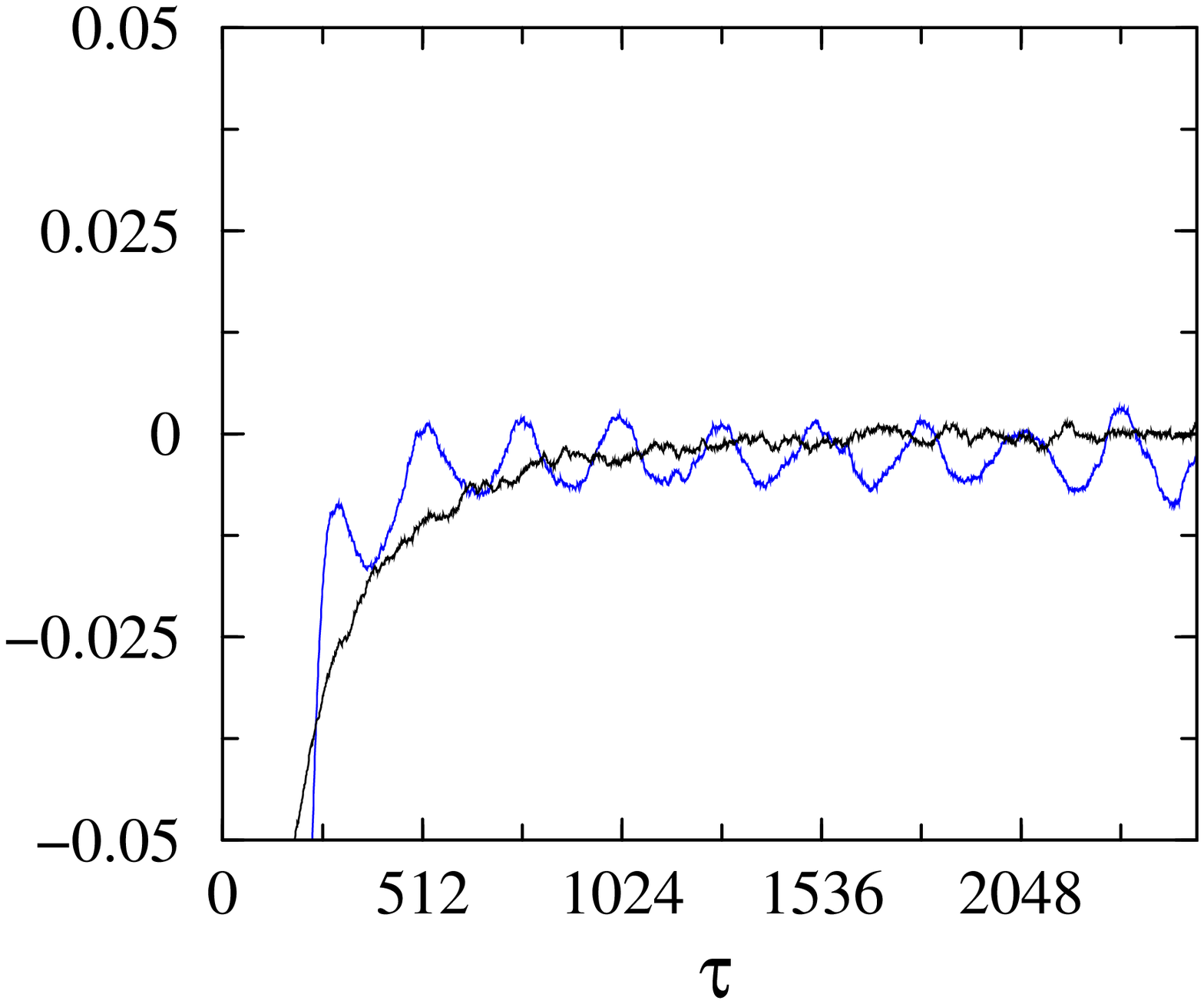}
}
\vspace{1em}
\caption{Correlation functions for the on-line game with real market histories. All data are from simulations at $P=\alpha N=128$, run for $10^5$ on-line time steps. Left column: $\rho=0.5$, the oscillatory curves are at $\alpha=0.1$, below the transition, non-oscillatory curves at $\alpha=1.25$, above the transition. Right column: $\rho=0.0$, oscillatory curves $\alpha=0.1$, non-oscillatory curves $\alpha=1.58$. The upper two panels show the `spin-spin' correlation function, $C_{ss}(\tau),$ conditional on the information pattern, the lower two panels the conditional correlation function, $C_{\pi\pi}(\tau)$, of the sign of the total bid $\pi(\ell)=\sgn[A(\ell)]$ (see text for further explanantions). Averages over $10$ to $100$ realisations of the disorder are taken. Results for $C_{\pi\pi}$ at $\rho=0$ are displayed as running averages over $64$ points to smoothen the noisy raw data. \label{fig:online_oscillations}  }
\end{figure}

Fig. \ref{fig:online_oscillations} demonstrates that the games both
with uncorrelated strategies ($\rho=1/2$), and with anti-correlated
strategies ($\rho=0$), respectively, show anti-persistence in their
non-ergodic states, $\alpha<\alpha_c(\rho)$, but not above $\alpha_c$:
$C_{\pi\pi}$ and $C_{ss}$ exhibit oscillations of period $2P$ below
$\alpha_c$, but approach a constant value above the transition. In
particular the behaviour of the on-line game with anti-correla\-ted
strategies appears crucially different from its batch counterpart in
this respect: in the batch game with anti-correlated strategies we
find oscillatory behaviour for all values of $\alpha$, whereas in the
on-line case with real histories and sign-updates, they are found only
in the low-$\alpha$ phase.

Before turning to the next section, we would like to remark that the oscillations below $\alpha_c$, first reported in \cite{ChalMars99}, are generally not detected very easily in on-line games: we have tested several other variations of the model and observables and found that no oscillatory behaviour can be observed if (i) the linear update (\ref{update}) is used instead of the sign-update prescription, (ii) if unconditional correlation functions are considered instead of conditional ones or if (iii) real market histories are replaced by fake histories. 

While no persistent oscillations are found in unconditional
correlation functions of on-line games with random histories,
oscillations emerge gradually for $\alpha<\alpha_c(\rho=0.5)$ in the
case of uncorrelated strategies, and for all $\alpha$ in the
anti-correlated case, as the time lag $M$ between two successive
strategy updates is increased to approach the batch limit, see
Fig. \ref{fig:oscillationsdelta} (oscillation amplitude for on-line
games is not shown, but is indistinguishable from zero on the scale of
Fig. \ref{fig:oscillationsdelta}).

\subsection{Random timing of adaptation}\label{sec:randomtiming}

\begin{figure}
\resizebox{0.4\textwidth}{!}{%
  \includegraphics{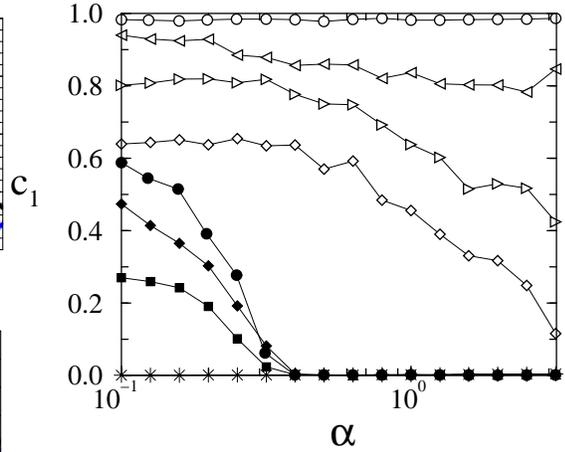} }
\vspace*{4mm} \caption{Amplitude of oscillations of the (unconditional) spin-spin correlation function of the on-line game with adaptation allowed only every $M$ time steps. Oscillations are of period $2M$.  All data is from simulations, the lines connecting the markers are guides to the eye. Solid symbols: $\rho=1/2$ (squares $M=10P$, diamonds $M=25P$, circles are the data for the batch limit,  plotted for comparison, see also Fig. \ref{fig:oscillations}). Open symbols: $\rho=0$ (diamonds $M=25P$, right triangles $M=100P$, left triangles $M=1000P$, circles are the data for the batch limit). All simulations are performed at $P=50$ and run for $10^6$ on-line steps, averages over $20$ realisations of the disorder are taken \cite{disclaimer}. Stars are the data for randomized updating ($\rho=0$, $M=10P$, see Fig. \ref{fig:volatility_1_m} for simulation details). No oscillations are found in this case.}
\label{fig:oscillationsdelta}
\end{figure}

\begin{figure}[t!!]
\resizebox{0.4\textwidth}{!}{%
  \hspace{1.5cm}\includegraphics{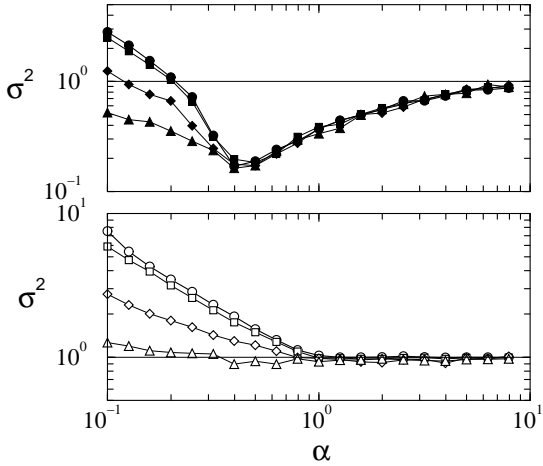}
}
\vspace*{4mm} \caption{Volatility of the game with randomized, asynchronous strategy updating (allow for strategy switches with probability $1/M$ at each on-line step, independently for different agents). Upper panel: uncorrelated strategies ($\rho=1/2$), lower panel: anti-correlated strategies ($\rho=0$). All data is from simulations at $P=50$, the top most curves (circles) are for $M=1$ (corresponding to the on-line game), squares $M=0.1P$, diamonds $M=P$, triangles $M=10P$. Simulations are started from {\em tabula rasa} initial conditions and run for $200M$ on-line steps ($2000M$ steps for $M=0.1P$), all data is averaged over at least $10$ realisations of the disorder. The horizontal lines mark the random trading limit, $\sigma^2=1$, in both panels \cite{disclaimer2}.
 }
\label{fig:volatility_1_m}
\end{figure}

Finally we have performed numerical simulations on MGs with
asynchronous, random timing of adaptation. Choosing $M\geq{\cal O}(P)$
and allowing all agents independently and randomly with probability
$1/M$ to update their strategy preferences at a given on-line step,
while still updating their score difference at each step, generates a
batch-like model with asynchronous updating. As depicted in
Fig. \ref{fig:oscillationsdelta} the randomization of the updates
removes the oscillations in the spin-spin correlation functions of the
batch game. The corresponding volatilities are virtually identical to
those of the corresponding on-line games above $\alpha_c$, in
particular we find $\sigma^2=1$ for $\alpha>1$ in the game with random
updating and full anti-correlation, see Fig. \ref{fig:volatility_1_m}.
In the non-ergodic phase the effect of the stochastic updating is a
gradual reduction of the volatility.
\section{Mixed population of speculators and producers}\label{sec:ps}

The theory of section \ref{sec:gfa} allows one in principle to study
the batch MG for an arbitray distribution $P(\rho)$ of correlation
parameters. While the previous section dealt only with the case of
unimodular distributions, we will now use the above formalism to study
a mixed population of `speculators' and `producers'
\cite{ChalMarsZhan00}. While a speculator is defined as a normal agent
(holding two strategies), a producer is an agent with limited choice
and has only one single strategy at his or her disposal (or
equivalently two identical strategies, corresponding to full
correlation, $\rho=1$). A detailed analysis of the statics of games
with such mixed populations is contained in \cite{ChalMarsZhan00}. In
this final section before our conclusions we will complement this work
by a study of the dynamics of such models, and will demonstrate that
the dynamical theory reproduces the results of the static replica
analysis. The interplay of producers and speculators is also discussed
in a different context in \cite{grandcangf}, where so-called
grand-canonical MGs are considered\footnote{In the grand-canonical
games studied in \cite{grandcangf} speculators are agents who hold
only one strategy (as opposed to two in the present paper), but in
addition they have the option not to play at a given
time step. Producers are agents with one strategy, but who play at
every time step.}; see also
\cite{MGbook2} for further details.

We will consider an ensemble of $N$ agents, consisting of $(1-x)N$
speculators (with correlation $0\leq\rho<1$ between their strategies)
and $xN$ producers, where $0\leq x < 1$. This corresponds to a choice
\be
P(\rho^\prime)=(1-x)\delta(\rho^\prime-\rho)+x\delta(\rho^\prime-1)
\ee
in the above generating functional calculation. As before, the
parameter $\alpha=P/N$ is defined as the ratio between the number of
patterns and the total number of agents. 
Using Eqs. (\ref{eq:ctot}) and (\ref{eq:chitot}) we find
\BE
c&=&\frac{1-\rho}{1-\overline{\rho}}(1-x)c(\rho)\\
\chi&=&\frac{1-\rho}{1-\overline{\rho}}(1-x)\chi(\rho), 
\EE
where $\overline{\rho}=(1-x)\rho+x$. $c$ and $\chi$ are then
determined self-consistently using expressions (\ref{gl:cofrho}) and
(\ref{gl:chiofrho}) for $c(\rho)$ and $\chi(\rho)$. We have checked
and confirmed these analytical results for $c$ against simulations for
some choices of the model parameters, but do not report the numerical
data here. After some more algebra we find
\be
\chi=\left(\frac{\alpha/(1-x)}{\erf\left(\lambda/\sqrt{2}\right)}-1\right)^{-1},
\ee  
where $\lambda$ is fixed as the root of the equation
\BE
&&\sqrt{\frac{2}{\pi}}\frac{e^{-\lambda^2/2}}{\lambda}+\left(1-\frac{1}{\lambda^2}\right)\erf\left(\frac{\lambda}{\sqrt{2}}\right)+\frac{\alpha/(1-x)}{\lambda^2}\nonumber\\
&=&\frac{1}{(1-\rho)(1-x)}.
\EE

\begin{figure}
\resizebox{0.38\textwidth}{!}{%
  \includegraphics{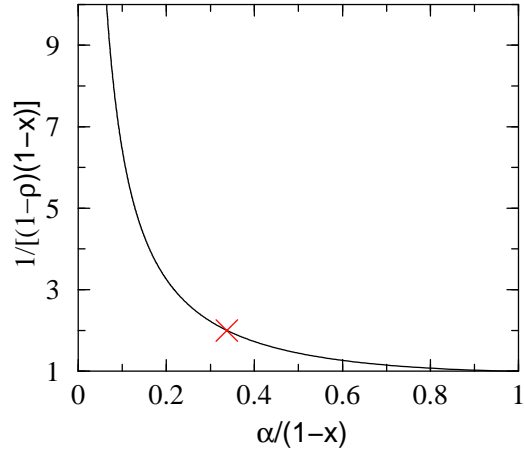}
}
\vspace*{4mm} \caption{Phase diagram of the batch MG with a mixed
population of speculators and producers. The line in the
$(\alpha/(1-x),1/[(1-x)(1-\rho)])$-plane is given by
(\ref{eq:criticallinemixedpop}). The cross marks the location of
the phase transition in the standard game $(q=0,\rho=0.5)$.}
\label{fig:phasediagram_with_producers}
\end{figure}

From this we locate the onset of diverging integrated response as
\be\label{eq:criticallinemixedpop}
\alpha_c=(1-x)\erf\left(\frac{\lambda_c}{\sqrt{2}}\right),
\ee
where $\lambda_c$ solves
\be
\sqrt{\frac{2}{\pi}}\frac{e^{-\lambda_c^2/2}}{\lambda_c}
+\erf\left(\frac{\lambda_c}{\sqrt{2}}\right)=\frac{1}{(1-\rho)(1-x)}.
\ee
Note that $\alpha_c/(1-x)$, depends only on the combination
$(1-\rho)(1-x)$. The resulting phase diagram is depicted in
Fig. \ref{fig:phasediagram_with_producers}. The above equations
obtained from the generating functional analysis agree with the
corresponding results from the replica calculation, as presented in
\cite{ChalMarsZhan00}\footnote{Note, however, that the conventions used in
\cite{ChalMarsZhan00} are slightly different, there an ensemble of $N$
speculators and $\rho N$ producers is considered (to make is
$(1+\rho)N$ agents in total), and $\alpha$ is defined as the ratio
between the number of patterns $P$ and the number of speculators,
$\alpha=P/N$.}.

Finally the generating functional approach allows one to study the
influence of decision noise on the phase diagram and behaviour of MGs
\cite{CoolHeimSher01}. A corresponding calculation for mixed
populations of producers and speculators shows that multiplicative
noise generally reduces the value of $\alpha_c/(1-x)$, i.e. that the range
of the ergodic phase is increased when the trading decisions of the
producers are made stochastically to a certain degree
\cite{GallaThesis}. This shift of $\alpha_c/(1-x)$ becomes more
pronounced as the noise level is increased. A similar effect was
previously also observed in MGs in which the agents trade on different
time scales \cite{DeMa03}.

\section{Concluding remarks}
We have presented an analysis of the dynamics of minority games with
diversified strategies. Generating functional techniques can be used
to turn the coupled dynamics of $N$ interacting agents with
heterogeneous strategy correlations into an effective single-particle
problem. The general case of heterogeneous correlation parameters
$\rho_i$ of the population of agents (drawn from a distribution
$P(\rho)$) is reflected in the fact that one finds an {\em ensemble}
of single agent processes as the final outcome of the theory, as
opposed to just one effective single-agent process for the case of
uniform correlation parameter ($\rho_i=\rho$ for all $i$). 

In section \ref{sec:ps} of this paper we have used the developed
formalism to study mixed populations of `speculators' and
`producers'. The dynamical approach leads to order parameter equations
which are identical to those obtained previously from replica analyses
of such models, and accordingly the phase diagrams obtained from the
statics and dynamics coincide. In general one finds that adding
producers to the MG increases the range of the ergodic phase of the
game.

The main focus of our study, however, has been the further analysis of
the dynamical effective single-agent problem for the cases of batch
and on-line games with unimodal distributions of the strategy
correlations, along with numerical simulations and complementing the
analysis of the statics of such games previously presented in
\cite{ChalMarsZhan00}.

We find that the model with uniform, but general correlation parameter
$\rho$ exhibits intriguing features, with similarities, as well as
crucial differences, compared to the game with uncorrelated strategies
($\rho=1/2$).

The main common features of the games with general correlation
parameter are (a) the existence of two distinct phases in both the
on-line and batch games for all $0\leq\rho<1$, with an ergodic state
for $\alpha>\alpha_c(\rho)$, in which no dependence on initial
conditions is found, and a non-ergodic phase below $\alpha_c(\rho)$,
in which the nature of the stationary state depends on the
configuration from which the dynamics is started; (b) persistent
oscillations are present in the non-ergodic phases of both the batch
and the on-line games for all $\rho$. Above $\alpha_c(\rho)$,
oscillations are absent in on-line games with arbitrary correlation
parameter $0\leq\rho<1$, and in batch games as long as $\rho>0$; (c)
for $0<\rho<1$ the transition point between the two phases is marked
by a minimum of the volatility $\sigma^2=\sigma^2(\alpha)$ in both
the batch and on-line games; (d) we find analytically that the
persistent order parameters in the stationary ergodic state and the
phase diagram do not depend on the details of the update rules (batch
versus on-line learning) and that they agree with those calculated
within the replica symmetric approximation; (e) no oscillations are
found in simulations of games with random asynchronous updating (for
neither uncorrelated nor anti-correlated strategies),
while at the same time a reduction of the volatility in the
non-ergodic phases of such games is observed.

However, the study of games with differently correlated strategies and
different timings of adaptation also reveals some new features and
striking differences between on-line and batch games, which up to now
have not been discussed systematically in the literature. These new
issues may be summarised as follows: (i) the dynamics of the batch
game with fully anti-correlated strategies, $\rho=0$, appears
different from the batch games with $0<\rho<1$ and from the on-line
game with $\rho=0$, as in the batch game at $\rho=0$ oscillations of
the form $C(\tau)=(-1)^\tau$ are found for all $\alpha$ and not only
in the low-$\alpha$ phase; for {\em tabula rasa} initial conditions
the volatility $\sigma^2$ of the batch game is a smooth function of
$\alpha$ without any minima or turning points. Nevertheless we find
that the game with fully anti-correlated strategies is in a
non-ergodic state for $\alpha<\alpha_c(\rho=0)=1$ for both batch and
on-line learning rules, with the usual dependence of macroscopic order
parameters on initial conditions in this regime; (ii) the volatilities
in batch and on-line games deviate increasingly from each other as the
correlation parameter $\rho$ is lowered; by allowing the agents to
update their strategy preferences synchronously only once every $M$
steps it is possible to interpolate smoothly between the on-line and
batch limits, $M=1$ and $M\geq {\cal O}(\alpha N)$, respectively;
(iii) the available approximations for the volatilities of batch MGs,
neglecting the retarded self-interaction of fickle agents, become
unreliable in this regime of anti-correlated strategy
assigments. Thus, care has to be taken whenever these approximations
are applied to other extensions of the conventional batch MG. The
corresponding approximations in the on-line case, based on discarding
transient contributions to the response and correlation functions,
however, appear to be valid above $\alpha_c(\rho)$ for all values of
$\rho$, even in the case of full anti-correlation.

\section*{Acknowledgements}

This work was supported by EPSRC (UK) under research grant GR/M04426
and studentship 00309273. TG acknowledges the award of a Rhodes
Scholarship and support by Balliol College, Oxford. Financial support
by the ESF programme SPHINX and by the European Community's Human
Potential Programme under contract HPRN-CT-2002-00319, STIPCO is
gratefully acknowledged. The authors would like to thank A C C
Coolen, J P Garrahan, M Marsili, E Moro, G Mosetti, M K
Y Wong and Y C Zhang for helpful discussions.


\begin{thebibliography}{99}
\bibitem{disclaimer}  We measure $c_1$ in the stationary state as $c_1=(2T)^{-1}\sum_{\tau=T}^{2T} |C(\tau+M)-C(\tau)|$, where $T$ corresponds to $25$ effective updates, i.e. $T=25M$ on-line steps. While the data for the uncorrelated case ($\rho=1/2$) does not seem to exhibit any strong dependence on the sample size $P$, the running time or the details of the numerical procedure of determining $c_1$ in the simulations, we find that the quantitative values of the data for the uncorrelated case ($\rho=0$) can depend on these parameters. This does not, however, affect the key qualitative observation of an increasing oscillation amplitude as $M$ is increased.
\bibitem{disclaimer2} The data for $\rho=0$ and $M=10P$ appears to lie (slightly) below the random trading limit $\sigma^2=1$ in an interval around $\alpha_c(\rho=0)=1$. This seems to persist also for larger values of $P$ or $M$ and/or for longer running times. It is at present unclear whether this constitutes a systematic effect or whether it is an artifact due to limitations in the numerical simulations. Note also that similar observations were made for the on-line case with real market histories ($\rho=0$ and $M=1$) \cite{ChalletThesis}.
\end{thebibliography}
\end{document}